\documentclass[onecolumn,10pt]{article}
\usepackage{amsmath}
\usepackage{amssymb}
\usepackage{amsfonts}
\usepackage{graphicx}
\usepackage{a4wide,epic}

\newcommand{\q}[1]{\vert #1 \rangle}
\newcommand{\qd}[1]{\langle #1 \vert}

\newcommand{\daag}{^{\dagger}}
\newcommand{\minou}{\text{-}}
\newcommand{\ba}{\text{\bf{a}}}
\newcommand{\bb}{\text{\bf{b}}}
\newcommand{\bn}{\text{\bf{N}}}
\newcommand{\bid}{\text{\bf{I}}}
\newcommand{\bnp}{{\bn+\bid}}

\newcommand{\uat}{\q{u_\text{at}}}
\newcommand{\bU}{\text{\bf{U}}}
\newcommand{\bbU}{\overline{\text{\bf{U}}}}

\newcommand{\bket}[1]{\langle #1 \rangle}

\newcommand{\rhoK}{\rho'^h}

\newcommand{\bbZ}{\overline{\bf{Z}}}

\begin{document}
\title{Stabilization of nonclassical states of one- and two-mode radiation fields by reservoir engineering}

\author{A. Sarlette\footnote{alain.sarlette@ugent.be (SYSTeMS, Ghent University, Technologiepark 914, 9052 Zwijnaarde, Belgium.)}\\ 
Z. Leghtas\footnote{(INRIA Paris-Rocquencourt, Domaine de Voluceau, Rocquencourt B.P.~105, 78153 Le Chesnay Cedex, France.)}\\ 
M. Brune\footnote{(Laboratoire Kastler-Brossel, ENS, UMPC-Paris6, CNRS, 24 rue Lhomond, 75005 Paris, France.)}\\ 
J.M. Raimond\footnote{(Laboratoire Kastler-Brossel, ENS, UMPC-Paris6, CNRS, 24 rue Lhomond, 75005 Paris, France.)}\\ 
P. Rouchon\footnote{(Centre Automatique et Syst\`{e}mes, Mines ParisTech, 60 boulevard Saint Michel, 75006 Paris, France)}}

\date{\today}

\maketitle


\paragraph*{{\large Abstract:}} We analyze a quantum reservoir engineering method, originally introduced by Sarlette \textsl{et al} [Phys. Rev. Lett. \textbf{107}, 010402 (2011)], for the stabilization of non-classical field states in high quality cavities. We generalize the method to the protection of mesoscopic entangled field states shared by two non-degenerate field modes. The reservoir is made up of a stream of atoms undergoing successive composite interactions with the cavity, each combining resonant with non-resonant parts. We get a detailed insight into the competition between the engineered reservoir and decoherence. We show that the operation is quite insensitive to experimental imperfections and that it could thus be implemented in the near future, either in the context of microwave Cavity Quantum Electrodynamics or in that of circuit-QED.\\

\maketitle



\section{Introduction}\label{sec:Intro}

Nonclassical electromagnetic field states are extremely important, both for a fundamental understanding of the quantum properties of light and for their possible use in practical applications. For instance, squeezed states (SS) have fluctuations of one of their quadratures below those of the vacuum state, or of a classical coherent state~\cite{Squeezing}. They lead thus to interesting methods for high-precision measurements and metrology~\cite{Giovannetti-al-Science_2004}. They are for instance planned to be used for  reducing the noise of the gravitational wave interferometers below the standard quantum limit~\cite{Goda08}.

Mesoscopic field state superpositions (MFSS) are also the focus of an intense interest. They involve a quantum superposition of two quasi-classical coherent components with different complex amplitudes. These counter-intuitive states bridge the gap between the quantum and the classical worlds and shed light onto the decoherence process responsible for the conspicuous lack of superpositions at our scale~\cite{HarocheBook}.

Finally, entangled superpositions of mesoscopic states (ESMS) shared by several field modes are even more intriguing. They violate generalized Bell inequalities~\cite{Banaszek-Wodkiewicz-PRL_1999}, illustrating the fundamentally non-local nature of quantum phyics. However, their non-local character is rapidly erased by a fast decoherence process~\cite{Milman-al-EurPhysJD_2005}, driving them back into a statistical mixture that can be undestood in terms of a classical local hidden variable model. This interplay of decoherence and nonlocality opens fascinating perspectives for exploring the limits of the quantum.

In principle, the SS and MFSS could be simply prepared in the optical domain by letting a coherent laser pulse propagate in a non-linear medium, whose index of refraction is a linear function of the light pulse intensity (Kerr medium)~\cite{YurkeStoler86}. The field evolves from initial coherent state $\q{\alpha}$ under the action of the Kerr Hamiltonian ${\bf H}_K$:
\begin{equation}\label{eq:Hkerr}
{\bf H}_{K} = \zeta_K \, \text{\bf{N}} \, +\, \gamma_K \, \text{\bf{N}}^2\, .
\end{equation}
Here $\text{\bf{N}}$ is the photon number operator, $\zeta_K $ is proportional to the linear index and $\gamma_K$ is the Kerr frequency describing the strength of the non-linearity. In the following, we use units such that $\hbar=1$. Note that the collisional interaction Hamiltonian for an $\text{\bf{N}}$-atom sample in a tightly confining potential or in an optical lattice is similar to ${\bf H}_K$~\cite{Blochcat}.

\begin{figure}
	\begin{center}
	\includegraphics[height=70mm,trim=0cm 0cm 0cm 0cm,clip=true]{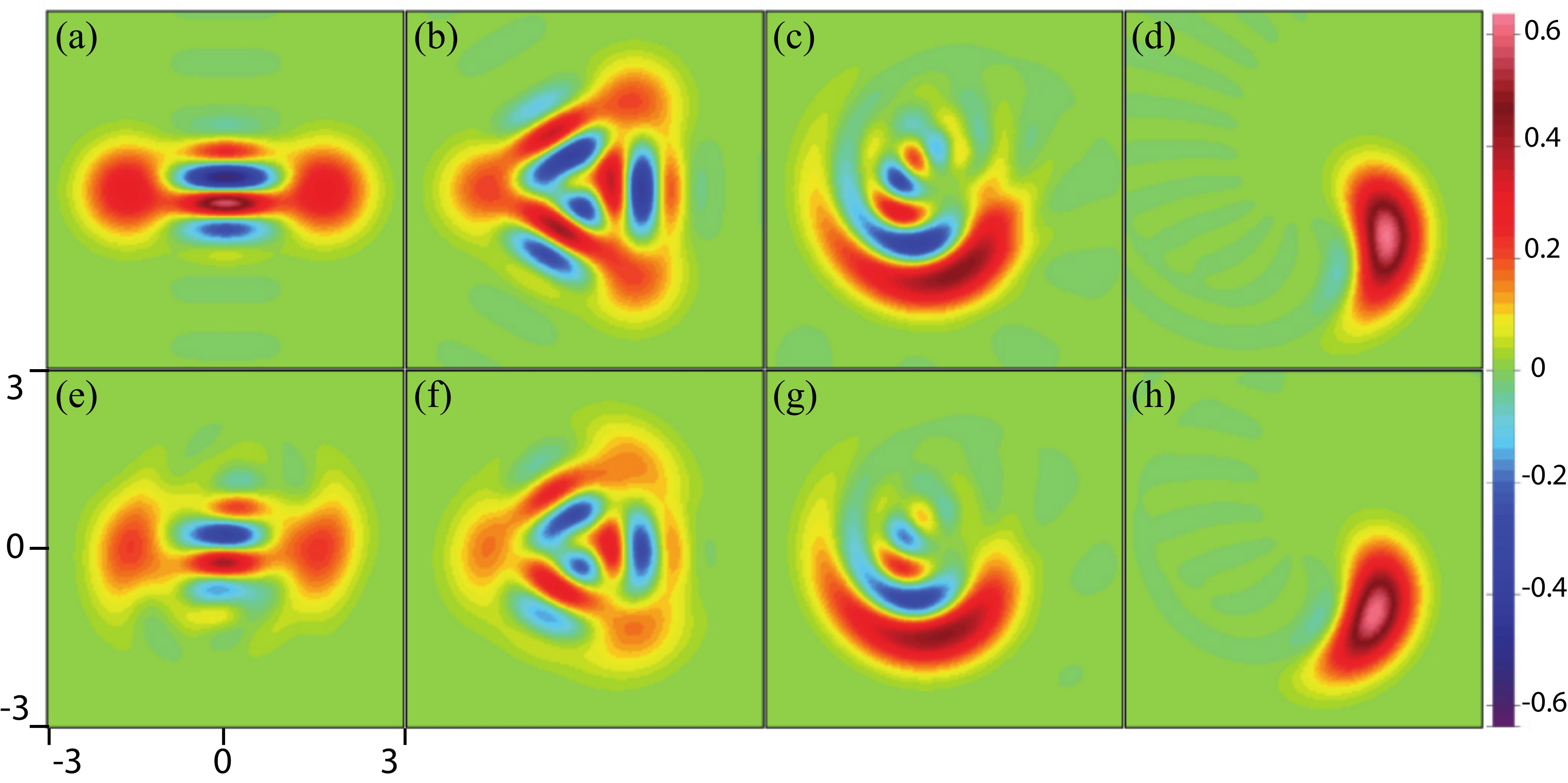}
\end{center}
	\caption{(a-d) Wigner functions of nonclassical field states $e^{-i\,t_K {\bf H}_K} \vert \alpha \rangle$ generated by propagation of an initial coherent state through a Kerr medium, (a) 2-component MFSS given by Eq.~\eqref{eq:forZak:CatOne} for $t_K \gamma_K=\pi/2$; (b) 3-component MFSS for $t_K \gamma_K=\pi/3$; (c) `banana'-state, for $t_K \gamma_K=0.28$; and (d) squeezed state, for $t_K \gamma_K =0.08\ll \pi$. (e-h): similar states stabilized, despite decoherence, by the atomic reservoir as explained in the remainder of the paper. Frame (e) corresponds to the reference two-component MFSS most lengthily discussed in the rest of the paper.}\label{fig:Hkerreds}
\end{figure}

Depending on the interaction time $t_K$, the final state $e^{-i\,t_K {\bf H}_K} \vert \alpha \rangle$ spans a number of nonclassical forms~\cite[Section~7.2]{HarocheBook}, including:
\begin{itemize}
\item[(i)] squeezed states for $t_K \gamma_K \ll \pi$;
\item[(ii)] states with `banana'-shaped Wigner function for slightly larger $t_K \gamma_K$;
\item[(iii)] mesoscopic field state superpositions $\q{k_\alpha}$ with $k$ equally spaced components for $t_K\gamma_K=\pi/k$~\cite{FNcoh}.
\item[(iv)] in particular, a MFSS of two coherent states with opposite amplitudes:
\begin{eqnarray}
\label{eq:forZak:CatOne} \q{c_{\tilde{\alpha}}} & = & ( \vert \tilde{\alpha} \rangle + i \, \vert \text{-}\tilde{\alpha} \rangle)/\sqrt 2\;,
\end{eqnarray}
with $\tilde{\alpha} = \alpha\, e^{-i \zeta_K t_K}$, for $t_K \gamma_K = \tfrac{\pi}{2}$.
\end{itemize}
The top panels on Figure \ref{fig:Hkerreds} present the Wigner functions of the states (i)-(iv) for a mean photon number $\vert \alpha \vert^2 = 2.7$.

This preparation method by a \emph{deterministic unitary evolution} is simple in its principle, but its implementation is extraordinarily difficult for propagating light fields, due to the weakness of the Kerr nonlinearity~\cite{Rosenbluh91}.

Other methods for the production of these nonclassical states have been proposed or realized in the context of trapped ions \cite{Winelandcat,Winelandres} or Cavity Quantum Electrodynamics (CQED)~\cite{HarocheBook,Brune92,Brune96,Deleglise08,Davidovich93,VillasB03,SolanoWalther03,deMatosFilho96}. Both systems implement the `spin-spring' model, the simplest nontrivial quantum situation of a two-level system coupled to a harmonic oscillator, embodied by the harmonic motion of the ion or by a single field mode. The proposed nonclassical state production methods are either deterministic or rely on a \emph{detection-conditioned} scheme. The latter expand the possibilities of the former by applying a measurement operation after a unitary evolution towards an intermediate target state. Measurement back-action generates different final states conditioned by the stochastic detection outcome~\cite{Grangier2007}. In the microwave CQED context, detection-conditioned preparation of MFSS and ESMS can be achieved by the dispersive interaction of an initial coherent field state with a two-level atom, initially prepared in a state superposition, followed by the detection of the atomic state in an appropriate basis~\cite{HarocheBook,LKB97}.

All these preparation techniques do not solve, however, the problem of \emph{stabilizing} (``protecting'') a selected nonclassical state for long times in spite of the unavoidable coupling of the system $\mathcal{S}$ to its environment $\mathcal{E}$. \emph{Reservoir engineering} can be used to stabilize target quantum states by strongly coupling $\mathcal{S}$ to an ``engineered'' environment, or reservoir  $\mathcal{R}$, a large quantum system with many degrees of freedom. The reservoir is designed so that, when acting alone, it drives  $\mathcal{S}$, whatever its intial state, towards a unique target `pointer state', a stable state of $\mathcal{S}$ coupled to $\mathcal{R}$~\cite{Zurek81,ZurekDeco}. The state of $\mathcal{S}$ remains close to this pointer state even in the presence of $\mathcal{E}$, provided  $\mathcal{R}$ is more strongly coupled to $\mathcal{S}$ than $\mathcal{E}$. An engineered reservoir thus achieves much more than the preparation of a target state. It effectively \emph{stabilizes} the system close to it for arbitrarily long times.

Reservoir engineering is experimentally challenging. Reservoirs made up of lasers and magnetic fields for trapped-ion oscillators have been proposed~\cite{Poyatos96,deMatosFilho96,Carvalho01} and demonstrated~\cite{Barreiro-et-al:Nature_2011}. Recently, a reservoir has been used to generate entanglement of spin states of macroscopic atomic ensembles~\cite{Krauter2011}.

In the context of CQED, the reservoir may be a stream of atoms interacting with the trapped field. An early proposal \cite{Meystre89} relied on the so-called `trapping state conditions' for the micromaser~\cite{Rempe-al-PRL_1990}, which require a very fine tuning of the parameters and can only be properly achieved in the case of a zero-temperature environment. Reservoirs composed of atoms in combination with external fields have also been proposed to stabilize one-mode squeezed states \cite{Werlang08} and two-mode squeezed vacuum entanglement~\cite{Pielawa07}.

In~\cite{PRancestor}, we proposed a robust reservoir engineering method for CQED. It generates and stabilizes nonclassical states of a single mode of the radiation field, including SS and MFSS. The reservoir is made up of a stream of 2-level atoms, each prepared in a coherent superposition of its basis states. They interact one at a time with the field according to the Jaynes-Cummings model before being discarded, a procedure reminiscent of the ``reset'' operation performed in other contexts \cite{Reed-al-Schoelkopf-APL_2010,Barreiro-et-al:Nature_2011}. The key idea is to use a tailored composite interaction of each atom with the field: dispersive, then resonant, then dispersive again. The pointer states of this composite interaction are precisely those, $e^{-i\,t_K {\bf H}_K} \vert \alpha \rangle$, resulting from the action of a Kerr Hamiltonian acting upon an initially coherent state.

This method is quite general and could be implemented in a variety of CQED settings, particularly in the active context of circuit QED~\cite{Devoret-Martinis-QInfoProcess_2004} or in that of microwave CQED, with circular Rydberg atoms and superconducting Fabry Perot cavities. For the sake of definiteness, we shall focus in this paper on the microwave CQED case, and particularly on the current ENS CQED experiment whose scheme is depicted on Fig.~\ref{fig:ExpLKB}.
The bottom panels of Fig.~\ref{fig:Hkerreds} present the results of numerical simulations of the ENS experiment, with interaction parameters chosen to reproduce the states generated by the Kerr Hamiltonian (top panels).

The present paper is intended to provide an in-depth description of this single-mode reservoir engineering procedure, with a detailed analysis of the physical mechanism of state stabilization. We discuss also the competition between the engineered reservoir and the ordinary cavity enviroment, giving simple insights into the finite final fidelity of the prepared state.

We finally extend the scheme proposed in \cite{PRancestor} to the stabilization of entangled superpositions of mesoscopic states of two field modes. The atoms of the reservoir undergo a tailored interaction with two modes of the same cavity, combining dispersive and resonant parts for each mode. This proposal opens interesting perspectives for studying the interplay between entanglement, non-locality and decoherence in the context of mesoscopic quantum states.

The paper is organized as follows. We consider the single-mode case for most of the paper and extend it to two modes in the last Section. Section \ref{sec:GenDes} describes the experimental scheme and the principle of the method. Section \ref{sec:Coherent} discusses, as a building block for the next Sections, how a stream of atoms resonant with one field mode stabilizes approximately a coherent field state. Section \ref{sec:LargeDetun} introduces the composite interaction: non-resonant, resonant and non-resonant again. In this Section, we treat the non-resonant interactions in the dispersive regime of a large atom-cavity detuning. We thus get a simple qualitative insight into the mechanism generating non-classical states. Section \ref{sec:ArbitDetun} details the more realistic case of intermediate atom-cavity detuning. We show that the main features of Section \ref{sec:LargeDetun} are recovered, exhibiting the robustness of the method. Section \ref{sec:decoherence} analyzes the effect of decoherence due to the cavity damping and imprecisions on the experimental parameters. We find that the method is also robust against realistically large imperfections. Section \ref{sec:2modes} finally presents the stabilization of two-mode ESMS.


\section{General description}\label{sec:GenDes}

\begin{figure}[!h]
	\begin{center}
\includegraphics[width=80mm]{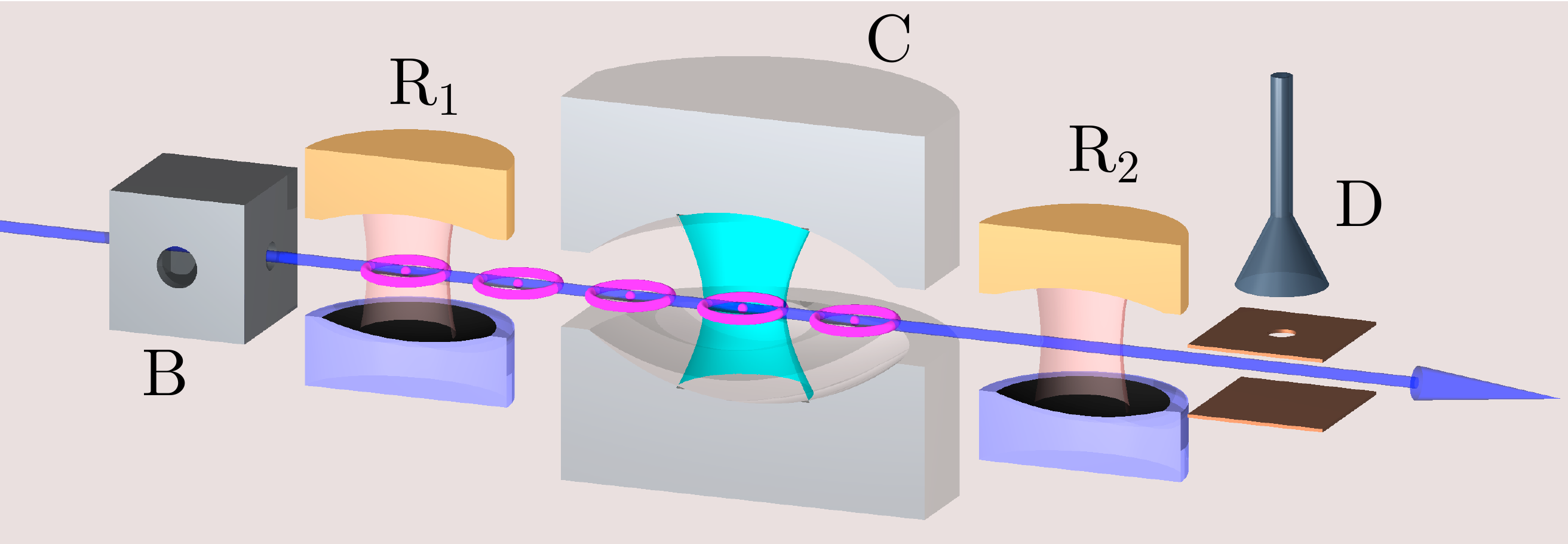}
	\end{center}
\caption{Scheme of the current ENS CQED experiment.}
\label{fig:ExpLKB}
\end{figure}

The scheme of the ENS experiment is depicted on Fig.~\ref{fig:ExpLKB} (see \cite{HarocheBook,Deleglise08} for details). A microwave field of frequency $\omega_c$ is trapped in the superconducting cavity $C$ (damping time $T_c=65$~ms). Atoms are sent one after the other through $C$. They cross its standing wave gaussian mode at a constant, adjustable velocity $v$. The mode interacts with the transition between the two atomic levels $\vert g \rangle$ and $\vert e \rangle$ (circular Rydberg states with principal quantum numbers 50 and 51). A static electric field applied across the cavity mirrors is used to adjust the atomic transition frequency $\omega_0$ and hence the atom-cavity detuning $\delta=\omega_0-\omega_c \ll \omega_c$ via the Stark effect. Varying the electric field during the atom-field interaction makes it possible to engineer the detuning profile $\delta(t)$. Zero and small $\delta$ values are used for the resonant and non-resonant parts of the interaction. Making $\delta$ very large allows us to effectively turn off the atom-field interaction.

We describe the atom and field states in a frame rotating at frequency $\omega_c$.
The atoms are prepared in state $\vert g \rangle$ in $B$, by a time-resolved laser and radiofrequency excitation of a velocity-selected thermal rubidium atomic beam. Before entering $C$, the atoms are prepared in a coherent superposition of $\q{g}$ and $\q{e}$ in the low-quality cavity $R_1$, driven by a classical microwave source (``first Ramsey zone'') at frequency $\omega_c$. Without loss of generality, we can choose the phase reference for all atoms so that they enter the cavity in the initial state $\vert u_\text{at} \rangle = \cos(u/2) \vert g \rangle +  \sin(u/2) \vert e \rangle$ with $u>0$. In a Bloch sphere representation with $\vert e \rangle$ at the north pole, $\vert u_\text{at} \rangle$ corresponds to a vector at an angle $u$ with the north-south vertical axis.

A second classical microwave pulse in the second Ramsey zone $R_2$ is followed by a detection in the $\{|e\rangle,|g\rangle\}$ basis in the field-ionization detector $D$. This amounts to a projective measurement of the atomic state at the exit of $C$, in a basis that can be chosen arbitrarily by properly setting the microwave pulse in $R_2$. For the engineered reservoir operation, the result of this final atomic state detection is irrelevant. Detection results are however necessary in other phases of the experiment. In particular, they will be used to reconstruct the field state generated by the engineered reservoir, using a method described in~\cite{Deleglise08}.

Let us first consider atom-cavity interaction for an atom that crosses cavity axis at $t=0$. The atom-field interaction is ruled by the Jaynes-Cummings Hamiltonian ${\bf H}_{JC}$. Neglecting far off-resonant terms (rotating wave approximation, negligible interaction with other cavity modes), we have:
\begin{equation}\label{eq:JCm}
  {\bf H}_{JC}= \frac{\delta(t)}{2}(\vert e \rangle \langle e \vert-\vert g \rangle\langle g \vert) + i\frac{\Omega(s)}{2} (\, \vert g \rangle \langle e \vert\, \text{\bf{a}}^{\dagger} - \vert e \rangle \langle g \vert\,  \text{\bf{a}} \, )\;,
\end{equation}
where $\Omega(s)$ is the atom-cavity coupling strength (vacuum Rabi frequency) at position $s=vt$ along the atomic trajectory; $\text{\bf{a}}$ is the photon annihilation operator in $C$. The photon number operator $\bn = \ba\daag\ba = \sum_n \; n \; \q{n} \qd{n}$ defines the Fock states basis $\{ \q{n} \}$.

The coupling strength $\Omega(s)$ is determined by the atomic transtion parameters and by the cavity mode geometry. It writes here $\Omega(s) = \Omega_0\; e^{-s^2 /w^2}$, with $\Omega_0/2\pi=50$~kHz and $w=6$~mm for the ENS setup. To get a finite total interaction duration $T$, we assume that the coupling cancels when $|s|>1.5\,w$. The total interaction time of the atom with the field is thus $T=3w/v$. We have checked numerically that this truncation of interaction time has negligible effects.

The evolution operator, or propagator ${\bf U}$ associated to ${\bf H}_{JC}$ expresses the transformation that the joint atom-cavity state undergoes during interaction. The Schr\"odinger equation for ${\bf U}$, starting at the initial time $t=t_0$ is:
\begin{equation}\label{eq:Uev}
	\tfrac{d}{dt} {\bf U}(t) = -i \, {\bf H}_{JC}(t) \, {\bf U}(t) \;\;\text{ with }\; {\bf U}(t_0)=\bid \, ,
\end{equation}
where $\bid$ is the identity operator. We note $\bU_T$ the propagator obtained by integration of Eq.~(\ref{eq:Uev}) over one full atom-cavity interaction, that lasts from $-T/2$ to $T/2$.

We represent the action of $\bU_T$ over the field state by the operators ${\bf M}^{\bU_T}_g$ and ${\bf M}^{\bU_T}_e$, such that:
\begin{eqnarray*}
\bU_T \; (\uat\q{\psi}) & = & \q{g}\, {\bf M}^{\bU_T}_g \q{\psi} + \q{e}\, {\bf M}^{\bU_T}_e \q{\psi}\;,
\end{eqnarray*}
for any pure initial field state $\q{\psi}$. Tracing over the final atomic state, the modification of the field density operator due to the interaction with the atom is thus finally given by the Kraus map~\cite{Kraus83}
\begin{equation}\label{eq:Kraus0}
\rho \rightarrow {\bf M}^{\bU_T}_g \rho {\bf M}^{{\bU_T}\,\dagger}_g + {\bf M}^{\bU_T}_e \rho {\bf M}^{{\bU_T}\,\dagger}_e \; .
\end{equation}

For the reservoir action, we let a stream of atoms consecutively interact with the field and always use the same parameter set (detuning profile, atom velocity $v$ and initial state $\vert u_\text{at} \rangle = \cos(u/2) \vert g \rangle +  \sin(u/2) \vert e \rangle$). Thus each atom affects the field according to Eq.~\eqref{eq:Kraus0}. The interaction of $C$  with the $(k+1)$th atomic sample begins as soon as that with the $k$th sample ends. Successive atoms are thus separated by the time interval $T$. We denote by $\rho_k$ the cavity state just after interacting with the $k$th atom and tracing over its irrelevant final state. The field density operator $\rho_{k}$ is thus given by:
\begin{equation}\label{eq:Kraus}
\rho_{k} = {\bf M}^{\bU_T}_g \rho_{k\minou 1} {\bf M}^{{\bU_T}\,\dagger}_g + {\bf M}^{\bU_T}_e \rho_{k\minou 1} {\bf M}^{{\bU_T}\,\dagger}_e \; .
\end{equation}
We aim to stabilize a pure pointer state $\rho_\infty =  \q{\psi_\infty} \qd{\psi_\infty}$, which must be a fixed point of this Kraus map. The right-hand side of Eq.~\eqref{eq:Kraus}, with $\rho_{k\minou 1} = \rho_\infty$, is a statistical mixture of two pure states. It is then a pure state only if its two terms are proportional to each other. Thus, $\q{\psi_\infty}$ must be an eigenstate of both ${\bf M}^{\bU_T}_g$ and ${\bf M}^{\bU_T}_e$. State stabilization by reservoir engineering amounts to tailoring a Kraus map for the field from a constrained physical setting.

We have shown in~\cite{PRancestor} that it is possible to engineer the atom-cavity interaction so that the Kraus map leaves invariant the states $\approx e^{-i \,t_K {\bf H}_{K}} \vert \alpha \rangle$, in which $\alpha$ and $\gamma_K\, t_K$ in Eq.\eqref{eq:Hkerr} can be chosen at will. Explicitly, we build $\bU_T$ by sandwiching a resonant interaction ($\delta=0$ for $t \in [-t_r/2,t_r/2]$) symmetrically between two dispersive interactions with opposite detuning: $\delta=\delta_0$ before the resonant interaction, $\delta=-\delta_0$ thereafter. We choose a positive $\delta_0$ value for the sake of definiteness.

\begin{figure}
		\begin{center}
\setlength{\unitlength}{1mm}
\begin{picture}(80,85)
\put(2,66){\includegraphics[width=80mm]{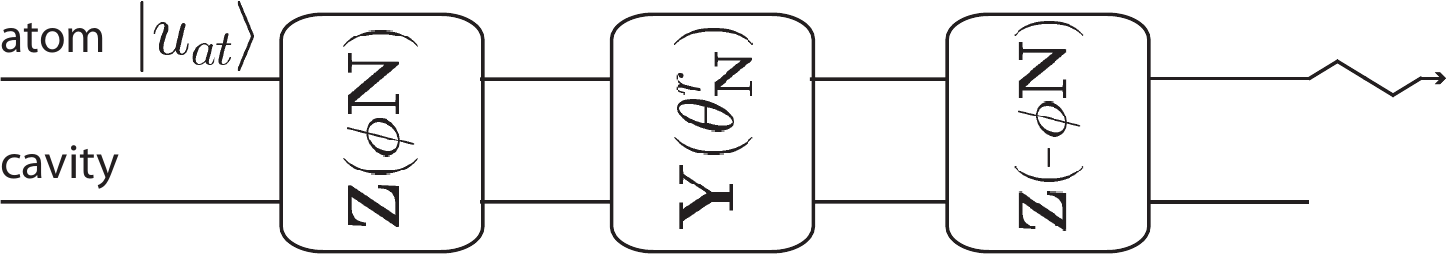}}
\put(0,0){\includegraphics[width=80mm]{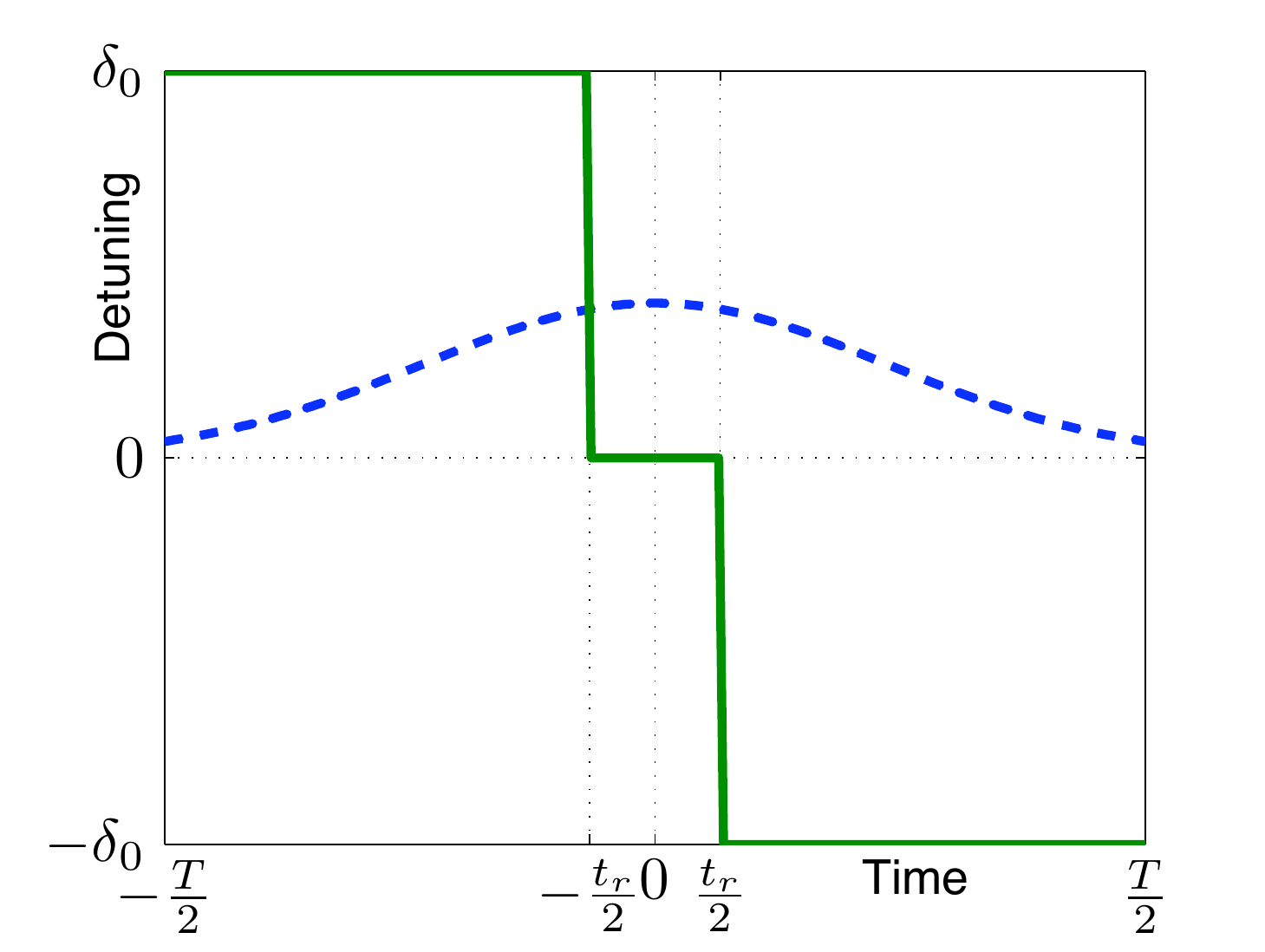}}
\end{picture}
	\end{center}
\caption{Time profile of $\delta$ (difference between the tunable frequency $\omega_0$ of the atomic transition and the fixed frequency $\omega_c$ of the single cavity mode; bottom frame, full line) and $\Omega(vt)$ (bottom frame, dashed line) during cavity crossing by one atomic sample; we take $t=0$ when the atom is at cavity center. Top frame: scheme of the propagators corresponding to the successive steps in the composite interaction.}\label{fig:timeline}
\end{figure}

This timing is illustrated on Fig.~\ref{fig:timeline}. Each resonant or dispersive interaction phase is characterized by a set of parameters that we denote $q=(t_1,t_2,v,\delta_0)$ where $t_1$ is start time, $t_2$ is stop time. The corresponding propagators are denoted ${\bf U}_q$.

To compute these evolution operators, we use the fact that each two-dimensional subspace spanned by $(\q{g,\,n+1},\q{e,\,n})$ is invariant under the action of ${\bf H}_{JC}$. The state $\q{g,0}$ does not take part in the evolution. We can thus view the action of $\bU_q$ as photon-number-dependent rotations acting  on a set of Bloch spheres $B_n$ ($n=0,1,...$), with $\q{g,n+1}$ at the south-pole and $\q{e,n}$ at the north-pole defining their $Z$-axes. These rotations can be decomposed as general rotations around the  $X$-, $Y$- and $Z$-axes of the Bloch spheres. We use the notation $f_{\bn} = f(\bn) = \sum_n \; f(n) \; \q{n}\qd{n} =  \sum_n \; f_n \; \q{n}\qd{n}$ for arbitrary functions $f$ of $n$, with the fundamental property
\begin{equation}\label{eq:NcommA}
	\ba \, f(\bn) = f(\bnp) \, \ba \; .
\end{equation}
We can then cast these rotations as:
\begin{eqnarray}
\label{eq:X}   {\bf X}(f_{\bn}) & = & \q{g}\qd{g} \, \cos (f_{\bn}/2) + \q{e}\qd{e} \, \cos (f_{\bnp}/2)\\
\nonumber & & -i \q{e}\qd{g} \, \ba \frac{\sin (f_{\bn}/2)}{\sqrt{\bn}} \, - i \q{g}\qd{e} \,  \frac{\sin (f_{\bn}/2)}{\sqrt{\bn}} \, \ba\daag \; ,\\
\label{eq:Y}   {\bf Y}(f_{\bn}) & = & \q{g}\qd{g} \, \cos (f_{\bn}/2) + \q{e}\qd{e} \, \cos (f_{\bnp}/2)\\
\nonumber & & - \q{e}\qd{g} \, \ba \frac{\sin (f_{\bn}/2)}{\sqrt{\bn}} \, + \q{g}\qd{e} \,  \frac{\sin (f_{\bn}/2)}{\sqrt{\bn}} \, \ba\daag \; ,\\
\label{eq:Z}   {\bf Z}(f_{\bn}) & = & \q{g}\qd{g}\;  e^{i\, f_\bn\; /2} + \q{e}\qd{e} \; e^{-i\, f_{\bnp}\; /2} \; .
\end{eqnarray}
As shown in Sections \ref{sec:Coherent} and \ref{sec:LargeDetun}, ${\bf Y}(f_{\bn}) $ with $f(n)$ proportional to $\sqrt{n+1}$ corresponds to a resonant interaction and ${\bf Z}(f_{\bn}) $ with $f(n)$ proportional to $n$ corresponds to a non-resonant interaction in the dispersive regime $\delta \gg \Omega$.
See Appendix~\ref{sec:propags} for more details.


\section{Engineered reservoir for coherent state stabilization}\label{sec:Coherent}

The coherent state $\q{\alpha}$ is   obtained by the action of the displacement operator ${\bf D}_{\alpha} = e^{\alpha \, \ba\daag- \alpha\daag \, \ba}$ onto the vacuum~\cite{HarocheBook}:
\begin{equation}\label{eq:CoherentState}
\q{\alpha} = {\bf D}_{\alpha} \, \q{0} \; = e^{-\vert \alpha \vert^2/2} \sum_n \, \frac{\alpha^n}{\sqrt{n!}} \q{n} \, .
\end{equation}
We show here how a short resonant interaction ($\delta=0$) with weakly excited atoms provides an engineered reservoir, whose pointer state is close to a coherent state \cite{PRancestor}.

Stabilization of coherent states is not an amazing feat. A classical radiation source weakly coupled to the cavity directly generates a coherent state. This is a routine operation in microwave CQED experiments. However, the situation described in this Section is an essential building block for the stabilization of more complex nonclassical states, as shown below. Moreover, it is an interesting micromaser situation~\cite{Meystre89,Casagrande02}, in which the small excitation of the atoms leads to a finite energy in the steady state even though the cavity is assumed to be lossless.

We consider a resonant interaction over a time interval $t_r$, corresponding to the parameter set $r=(-t_r/2,t_r/2,v,0)$. Following Appendix~\ref{sec:propags}, the associated propagator is:
\begin{eqnarray}
\label{eq:Uresonant} 	
 {\bf U}_r \,\, & = \,\,\, & {\bf Y}(\theta^{r}_{\bn}) \\
\nonumber
	 & = \,\,\, & \q{g}\qd{g} \, \cos (\theta^{r}_{\bn}/2) + \q{e}\qd{e} \, \cos (\theta^{r}_{\bnp}/2)\\
\nonumber
\!\! & & - \q{e}\qd{g} \, \ba \frac{\sin (\theta^{r}_{\bn}/2)}{\sqrt{\bn}} \, + \q{g}\qd{e} \,  \frac{\sin (\theta^{r}_{\bn}/2)}{\sqrt{\bn}} \, \ba\daag\;,
\end{eqnarray}
with
\begin{equation}\label{eq:thetaR}
\theta^{r}_n = \theta_r \sqrt{n}, \quad \theta_r = \int_{-t_r/2}^{t_r/2} \, \Omega(v t) \, dt \; .
\end{equation}
This readily yields:
\begin{eqnarray}
\nonumber
{\bf M}^{{\bf U}_r}_g & = & \cos(\tfrac{u}{2}) \, \cos (\theta^{r}_{\bn}/2) + \sin(\tfrac{u}{2}) \, \frac{\sin (\theta^{r}_{\bn}/2)}{\sqrt{\bn}} \, \ba\daag\\
\label{eq:res:MgMe}	
{\bf M}^{{\bf U}_r}_e & = & \sin(\tfrac{u}{2}) \, \cos (\theta^{r}_{\bnp}/2) - \cos(\tfrac{u}{2}) \, \ba \frac{\sin (\theta^{r}_{\bn}/2)}{\sqrt{\bn}} \; . \phantom{kkk}
\end{eqnarray}

A pointer state of this resonant reservoir must be an eigenstate of both ${\bf M}_g$ and ${\bf M}_e$ given by Eq.(\ref{eq:res:MgMe}). Let us expand it over the Fock states basis, $\q{\psi_\infty} = \sum \, \psi_n \q{n}$. We get a condition on the coefficients $\psi_n$, for $n=0,1,2,...$:
\begin{equation}\label{eq:res:pscond}
\sin (\theta^{r}_{n+1}/2) \, \psi_{n+1} \; = \; \tan \tfrac{u}{2} \, (1+\cos (\theta^{r}_{n+1}/2)\,) \, \psi_n \, .
\end{equation}
This relation allows to compute all the $\psi_n$ starting from $\psi_0 \neq 0$, except if $\sin (\theta^{r}_m/2) = 0$ for some $m$. This condition corresponds to the existence of a trapping state $\q{m-1}$~\cite{TrappingConditions}, which is then uncoupled from $\q{m}$. The Hilbert subspaces corresponding to the photon numbers $\leq (m-1)$ and to those $\geq m$ are then decoupled, such that the steady state depends on the initial conditions. Since all states considered in the remainder of the paper have an energy lower than $20$ photons, we arbitrarily truncate the Hilbert space to $n \leq n_{\max} = 50$. We can thus avoid trapping states by choosing small $\theta_r$ values such that $\sin (\theta^{r}_{n+1}/2) \neq 0$ for all $0 \leq n \leq n_{\max}$. Dividing (\ref{eq:res:pscond}) by $\sin (\theta^{r}_{n+1}/2)$ then gives the recurrence:
\begin{equation}\label{eq:res:iter}
\psi_{n+1} \; = \; \frac{\tan (u/2)}{\tan (\theta^{r}_{n+1}/4)} \, \psi_n \ ,
\end{equation}
which defines a unique normalized pointer state.

For $(\theta^{r}_{n_{\max}}/4)^2 \ll 1$, the recurrence (\ref{eq:res:iter}) reduces to $\psi_{n+1} \; \approx \; \frac{4\tan(u/2)}{\theta_r \sqrt{n+1}} \, \psi_n$, which defines a coherent state $\q{\alpha_\infty}$ with $\alpha_\infty = \tfrac{4\tan(u/2)}{\theta_r}$ (compare with the last member of Eq.~\eqref{eq:CoherentState}).

This value of $\alpha_\infty$ can be retrieved by a simplified reasoning as in~\cite{PRancestor}. Assume that the cavity already contains a large coherent field of amplitude $\alpha\gg 1$. The incoming atoms then undergo a resonant Rabi rotation in this field, with an atomic Bloch vector starting initially towards the south pole of the Bloch sphere. The Bloch vector rotates by an angle $-\theta_r \alpha$, such that  if $\theta_r\alpha<2u$ (resp.~$\theta_r \alpha >2u$) the final atomic state has a lower (resp.~larger) energy than the initial one, i.e.~in average gives energy to (resp.~draws energy from) the field. This energy exchange thus stabilizes a field with amplitude $\alpha_\infty = 2u/\theta_r$.

\begin{figure}
	\begin{center}
\includegraphics[width=80mm, trim=0cm 0cm 9cm 1.2cm,clip=true]{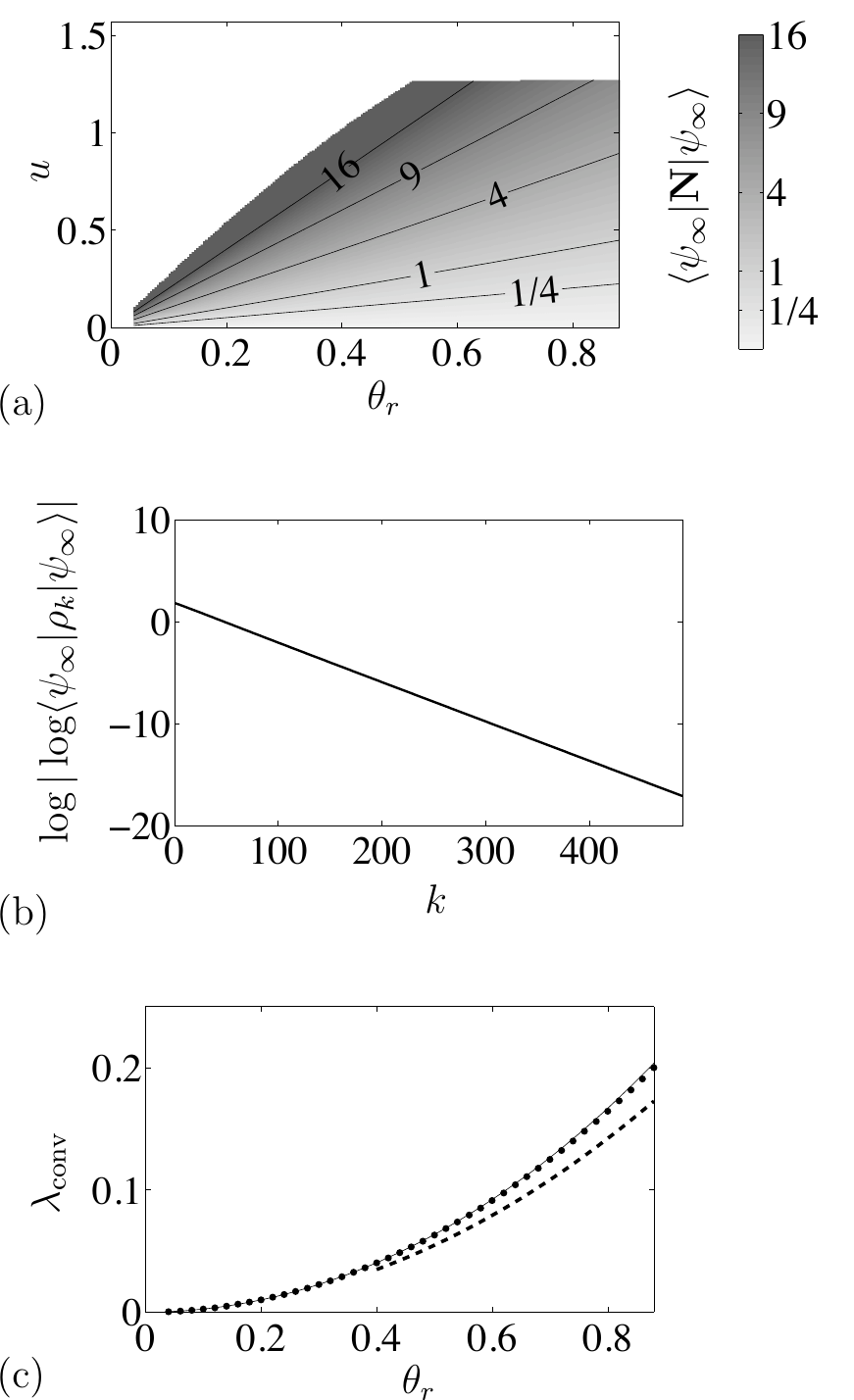}
\end{center}
\caption{Reservoir with interaction $\bU_r$. (a) Mean photon number $\qd{\psi_\infty} \bn \q{\psi_\infty}$ of the pointer state $\q{\psi_\infty}$. Grayscale axis is linear in $\sqrt{\qd{\psi_\infty} \bn \q{\psi_\infty}}$. The shaded zone is delimited such that the corresponding states have at least a $99\%$ fidelity $\vert \langle \psi_\infty \q{\alpha_*} \vert^2$ to a coherent state $\q{\alpha_*}$ of same mean photon number $\vert \alpha_* \vert^2 = \qd{\psi_\infty} \bn \q{\psi_\infty}$. On the top left corner, pointer states have significant population outside the truncated Hilbert space. On the top right part, $\vert \langle \psi_\infty \q{\alpha_*} \vert^2$ drops to $\sim98\%$ as $u$ approaches $\pi/2$. (b) Evolution of the fidelity indicator $\log \vert \log \qd{\psi_\infty} \rho_k \q{\psi_\infty}\vert$  as a function of the number of atom-field interactions (i.e.~Kraus map iterations) $k$, starting from vacuum $\rho_0 = \q{0}\qd{0}$. We have  arbitrarily set $u=0.5$ and $\theta_r=0.4$, for which $\qd{\psi_\infty} \bn \q{\psi_\infty} = 6.21$. (c) Convergence rate $\lambda_{\text{conv}}$ as a function of $\theta_r$ for $u=0.1$ (dotted curve) and $u=1$ (dashed curve). Dependency in $u$ is small. We also represent (full curve) the analytic result of the simplified model [Eq.~\eqref{eq:Zdyns}]. This model slightly overestimates the convergence speed.}\label{fig:CohEvMap}
\end{figure}

We have numerically examined the fidelity $F= \vert \qd{\psi_\infty} \alpha_* \rangle \vert^2$ of the pointer state $\q{\psi_\infty}$ defined by Eq.~(\ref{eq:res:iter}) with respect to a coherent state $\q{\alpha_*}$ of the same mean photon number $\vert \alpha_* \vert^2 = \qd{\psi_\infty} \bn \q{\psi_\infty}$. Figure \ref{fig:CohEvMap}(a) represents that mean photon number in gray scale, for a range of parameters $u,\theta_r$ delimited such that the fidelity $F$ is larger than 99\%. We limit the plot to $\;\;\theta_r < (2\pi) / \sqrt{n_{\max}} \approx 0.88$ to avoid trapping states, and to $\;\; \theta_r > 5\tan(u/2)/\sqrt{n_{\max}}$ to remain in the truncated Hilbert space (top left corner cut off).
The coherent state approximation for $\q{\psi_\infty}$ remarkably holds for a range of $u$ and $\theta_r$ much larger than that predictable from the qualitative discussion above.

Convergence towards $\q{\alpha_\infty}$ can be simply analyzed in the limit of small $u,\theta_r$. Expansion of Eq.~(\ref{eq:res:MgMe}) to second order in $u,\,\theta^{r}_\bn$ yields the Kraus map:
\begin{eqnarray}
\label{eq:KrausApprox}
\rho_{k+1} & \approx & \rho_k + \frac{u \theta_r}{4}\, (\, [\ba\daag,\rho_k] - [\ba,\rho_k] \,) \\
\nonumber & & - \frac{\theta_r^2}{8}\, (\bn \rho_k + \rho_k \bn - 2 \ba \rho_k \ba\daag)\ .
\end{eqnarray}
It can be simplified by letting $\tilde{\rho} = {\bf D}_{-\alpha_{\infty}} \, \rho \, {\bf D}_{\alpha_{\infty}}$ such that $\rho=\q{\alpha_\infty}\qd{\alpha_\infty}$ corresponds to $\tilde{\rho}=\q{0}\qd{0}$. A few calculations show that Eq.~(\ref{eq:KrausApprox}) transforms into
\begin{equation}\label{eq:GotoVac}
	\tilde{\rho}_{k+1} = \tilde{\rho}_k - \frac{\theta_r^2}{8} \, (\bn \tilde{\rho}_k + \tilde{\rho}_k \bn - 2 \ba \tilde{\rho}_k \ba\daag) \, .
\end{equation}
This is a finite difference version of the standard Lindlblad equation, describing the damping of an harmonic oscillator coupled to a zero temperature bath. It efficiently drives any initial state towards the vacuum, with a relaxation rate proportional to $\theta_r^2$. This analogy shows that the initial map [Eq.~(\ref{eq:KrausApprox})] drives any initial cavity state towards a coherent state $\q{\alpha_\infty}$ with $\alpha_\infty=2u/\theta_r$. Smaller $\theta_r$ values, i.e. shorter interaction times of each atom with the field lead to a higher energy pointer state (for a given $u$), but to a lower convergence rate (independently of $u$).

Similar conclusions are directly reached from Eq.~(\ref{eq:KrausApprox}) by assuming that the field is at any stage during its evolution towards equilibrium in a coherent state with amplitude $\alpha_k$. Using simple second-order approximations in $u,\theta_r$~\cite{PierreApprox}, we find that this amplitude evolves as:
\begin{equation}\label{eq:Zdyns}
	\alpha_{k+1} = (1-\theta_r^2 / 8) \alpha_k + u \theta_r / 4 \, .
\end{equation}
This first-order system has the explicit solution
$ \alpha_k = (1-\theta_r^2 / 8)^k \, (\alpha_0 - \alpha_\infty) + \alpha_\infty $
starting from $\alpha_0$ at $k=0$.
Noting that $\log \vert \langle{\alpha_\infty} \q{\alpha_k} \vert^2 = -\vert \alpha_k - \alpha_\infty \vert^2$, the fidelity indicator $\log \vert \log \vert \langle{\alpha_\infty} \q{\alpha_k} \vert^2 \vert = \log \vert \alpha_0 - \alpha_\infty \vert^2 - \lambda_{\text{conv}}\, k$ decreases linearly in $k$ towards $-\infty$.
The slope $\lambda_{\text{conv}} = 2 \vert \log(1-\theta_r^2/8) \vert$ measures the exponential convergence speed of $\vert \alpha_k - \alpha_\infty \vert^2$, which increases with $\theta_r$ and is independent of $u$.

Numerical simulations of Eq.~(\ref{eq:Kraus}) with the exact Kraus map [Eq.~(\ref{eq:res:MgMe})] vindicate this approximate analysis.
Figure~\ref{fig:CohEvMap}(b) shows the evolution of $\log \vert \log \qd{\psi_\infty} \rho_k \q{\psi_\infty} \vert$ as a function of the number of atom-field interactions  $k$, starting from the vacuum $\rho_0 = \q{0}\qd{0}$, with the real Kraus map associated to $\bU_r$. The evolution is linear, as predicted by the simplified model. We have checked that this linearity holds for a large range of parameter values: it is only for large $\theta_r$s that the curve bends slightly upwards for the first few $k$s. This allows us to use the slope $\lambda_{\text{conv}}$ of that approximate line as a measure of convergence speed. Fig.~\ref{fig:CohEvMap}(c) shows the dependency of $\lambda_{\text{conv}}$ in $\theta_r$, for two different $u$ values: $u=0.1$ (dotted curve) and $u=1$ [dashed curve, which does not extend to low $\theta_r$ values, according to the accessible domain on Fig.~\ref{fig:CohEvMap}(a)]. They closely follow the simplified model (full line), which is independent of $u$ and slightly overestimates convergence speed.

\section{Kerr Hamiltonian simulation in the dispersive regime}\label{sec:LargeDetun}

We now discuss the case of a full composite interaction, with the detuning profile $\delta(t)$ represented on Fig.~\ref{fig:timeline}.
The full propagator
\begin{equation}\label{eq:CompInt}
	\bU_T = \bU_{d_2} \; \bU_r \; \bU_{d_1}\;,
\end{equation}
is the concatenation of three unitary operators corresponding first to the dispersive interaction with $d_1=(-T/2,-t_r/2,v,\delta_0)$ as the atom enters the cavity, then to the resonant one with $q=r$ around cavity center, then to the last dispersive interaction with $d_2=(t_r/2,T/2,v,-\delta_0)$ as the atom leaves the cavity. The exact expression of $\bU_r$ is given by Eq.~\eqref{eq:Uresonant}. The dispersive propagators $\bU_{d_1}$ and $\bU_{d_2}$ are computed in Appendix~\ref{sec:propags}, assuming that $\delta_0$ and $v$ satisfy the adiabatic approximation condition [Eq.~\eqref{eq:AdCond}]. This condition merely expresses that the interaction Hamiltonian ${\bf H}_{JC}$ varies slowly (through $\Omega(vt)$) in comparison to the differences between its eigenfrequencies. Thus, each initial eigenstate of ${\bf H}_{JC}$ remains an eigenstate and only accumulates a phase proportional to its eigenenergy.

To get a simple insight in the physics of the stabilization of nonclassical states, the present Section focuses (like~\cite{PRancestor}) on the simple case in which the two nonresonant interactions take place in the dispersive regime, i.e. $\delta_0\gg \Omega_0$. This avoids
spurious population transfers in the dispersive phase, as atomic levels dressed by the cavity field almost coincide with the bare levels $\q{e}$ and $\q{g}$. The dispersive propagators, deduced from Eq.~\eqref{eq:Uadiabatic} in Appendix~\ref{sec:propags}, then write:
\begin{eqnarray}
\label{eq:DispIntZ}
{\bf U}_{d_1} & \approx & {\bf U}_{d_2}^\dag  \approx \,\,\, {\bf Z}(\phi_\bn^d)\\[2mm]
\nonumber \text{with } \phi_\bn^d & = & \phi_\gamma \bn + \phi_\zeta \; ,
\end{eqnarray}	
where $\phi_\gamma = 1/(2\, \delta_0)\, \int_{-T/2}^{-t_r/2} \Omega^2(vt)\,dt$ is a phase shift per photon and $\phi_\zeta = \delta_0 (T-t_r)/2$ reflects the free atom evolution in the interaction representation at cavity frequency.

The full propagator then writes
\begin{eqnarray}
\nonumber \bU_T  \approx \bU_d & = & {\bf Z}(-\phi_\bn^d)\;\; \bU_r \;\; {\bf Z}(\phi_\bn^d) \\
\nonumber & = & \q{g}\qd{g} \, \cos \theta^{r}_\bn/2 + \q{e}\qd{e} \, \cos \theta^{r}_{\bnp}/2\\
\nonumber & & - \q{e}\qd{g} \, \ba \, \frac{\sin \theta^{r}_\bn/2}{\sqrt{\bn}} \, e^{i (\phi_\gamma \bn + \phi_\zeta)}\\
\label{eq:Utsimp} & & + \q{g}\qd{e} \, \frac{\sin \theta^{r}_\bn/2}{\sqrt{\bn}} \, e^{-i (\phi_\gamma \bn + \phi_\zeta)} \, \ba\daag\;,
\end{eqnarray}
where $\theta^r_\bn$ is defined by Eq.~\eqref{eq:thetaR}. The opposite dispersive interactions have no net effect when the atom remains in the same state during the resonant interaction $\bU_r$. In contrast, ${\bf Z}$ does not commute with terms in which the atomic level changes in the resonant phase. For these terms, the dispersive phase shifts add up. The global evolution $\bU_d$ thus associates a phase shift to each term of $\bU_r$ that changes the field energy. An increasing field energy corresponds to a decrease of the field phase and vice versa. These correlated phase and amplitude shifts suggest that $\bU_d$ might stabilize a coherent state distorted by amplitude-dependent phase shifts,  a situation similar to that encountered during the propagation through a Kerr medium.

The action of the atom can indeed be expressed  by an operator acting on the field only. Let us define the Hermitian operator $h^d_{\bn}$ by:
\begin{equation}\label{eq:h0}
	 h^d_{\bn} = \phi_\gamma (\bn^2 + \bn)/2 + \phi_\zeta \bn \, .
\end{equation}
With the commutation identity [Eq.~(\ref{eq:NcommA})] we have  $\; e^{-i h^d_{\bn}} \, \ba \, e^{i h^d_{\bn}} = \ba \, e^{i (\phi_\gamma \bn + \phi_\zeta)} \;$ and
\begin{equation}\label{eq:CVs}
\bU_d = e^{-i h^d_{\bn}} \; \bU_r \; e^{i h^d_{\bn}} \; .
\end{equation}
Thus, $\bU_d$ is equivalent to $\bU_r$ modulo a basis change on field state alone defined by the unitary operator $e^{-i h^d_{\bn}}$.

In other words, when $\rho$ evolves under the Kraus map associated to $({\bf M}^{{\bf U}_d}_g\, , \; {\bf M}^{{\bf U}_d}_e)$,
$\rho^h= e^{i h^d_{\bn}} \; \rho \; e^{-i h^d_{\bn}}\; ,$
evolves under the Kraus map associated to $({\bf M}^{{\bf U}_r}_g\, , \; {\bf M}^{{\bf U}_r}_e)$. It follows from Section~\ref{sec:Coherent} that $\rho^h$ converges towards a coherent pointer state $\q{\alpha_\infty}$. Therefore, $\rho$ converges with the same convergence rate towards a nonclassical pointer state $\exp[-i {h}^d_{\bn}]\, \q{\alpha_\infty}$.

The effective Hamiltonian $h^d_{\bn}/t_K$ is equal to the Kerr Hamiltonian ${\bf H}_K$, with $\gamma_K t_K = \phi_\gamma/2$ and $\zeta_K t_K=(\phi_\zeta+\phi_\gamma/2)$. The engineered reservoir thus stabilizes the nonclassical pointer states $e^{-i t_K {\bf H}_K} \, \q{\alpha_\infty}$ which would be produced by propagation through a Kerr medium (see Fig.~\ref{fig:Hkerreds}). Tuning $T$ and $\delta_0$ allows us to choose $\phi_\gamma$ at will. We can thus prepare and stabilize a whole class of such states, as described in Section \ref{sec:Intro}. In particular, for $\phi_\gamma = \pi$, we get the MFSS $\vert c_{\tilde{\alpha}_\infty} \rangle=( \vert \tilde{\alpha}_\infty \rangle + i \, \vert \text{-}\tilde{\alpha}_\infty \rangle)/\sqrt 2$ with $\tilde{\alpha}_\infty =  e^{-i\,(\phi_\zeta+\pi/2)}\, \alpha_\infty$. Note that the stabilization of this two-component MFSS is the most demanding experimentally, since it requires the longest dispersive interaction time.

The discussions in this Section only apply in the limit of small $\Omega/\delta_0$. Reaching notable $\phi_\gamma$ in this case requires a large dispersive interaction time $(T-t_r)/2$, that can be prohibitive. First, in the experimental context of Fig.~\ref{fig:ExpLKB}, $T=3w/v$ is limited by the minimal achievable atomic velocity (a few tens of m/s in the ENS setup). Second, a larger $T$ means less frequent atom-field interaction and thus a weaker reservoir, implying a less efficient protection of the target state against decoherence induced by cavity relaxation.



\section{Regime of arbitrary detunings}\label{sec:ArbitDetun}

We thus discuss now dispersive interaction with moderate $\Omega/\delta_0$ values, which allows to reach significant dispersive effects within moderate interaction times. We therefore use a more precise expression of the propagator for the nonresonant interactions (parameter sets $d_1$ and $d_2$), by applying the adiabatic approximation to the actual dressed states (instead of $\q{g,n+1}$ and $\q{e,n}$ as in Eq.~\eqref{eq:DispIntZ} when assuming $\delta_0 \gg \Omega_0$). Developments detailed in Appendix~\ref{sec:propags} lead to:
\begin{equation}\label{eq:UcOps}
	\bU_T \approx \bU_{c} = {\bf Z}(-\phi_{\bn}) \, {\bf X}(\xi_{\bn}) \, {\bf Y}(\theta^r_{\bn}) \, {\bf X}(\xi_{\bn}) \, {\bf Z}(\phi_{\bn})\;,
\end{equation}	
with
\begin{eqnarray}
\label{eq:phiDelta}  \phi_n & = & \delta_0 \, \int_{-T/2}^{-t_r/2} \, \sqrt{1 + n\, (\Omega(vt)/\delta_0)^2}\; dt \ , \\
\label{eq:xiDelta}   \tan \xi_n & = & \tfrac{\Omega(vt_r/2) \sqrt{n}}{\delta_0} \quad \text{with } \xi_n \in (\tfrac{-\pi}{2},\tfrac{\pi}{2}) \; .
\end{eqnarray}
We recognize in this expression the central resonant interaction evolution operator, $ {\bf Y}(\theta^r_{\bn}) $, and the two phase-shift operations accumulated during the non-resonant interactions (${\bf Z}(-\phi_{\bn}) $ and ${\bf Z}(\phi_{\bn})$). Note that here, unlike in Section \ref{sec:LargeDetun}, $\phi_n$ is a nonlinear function of $n$.
The remaining two ${\bf X}(\xi_{\bn})$ operators reflect the fact that the atomic energy eigenstates do not coincide with the dressed levels at $\pm t_r/2$, when the atomic frequency is suddenly switched.
Note that we neglect two similar transformations which appear in principle when the atom gets first coupled to the mode and finally decoupled from it, since the atom-field coupling is then quite negligible.

Some tedious but simple computations exploiting Eq.~\eqref{eq:NcommA} allow us to write:
\begin{eqnarray}
\nonumber \bU_{c} & = & 	\q{g}\qd{g} \, \cos (\theta^c_{\bn}/2) + \q{e}\qd{e} \, \cos (\theta^c_{\bnp}/2)\\
\nonumber & &  - \q{e}\qd{g} \, \ba \frac{\sin (\theta^c_{\bn}/2)}{\sqrt{\bn}} \; e^{i \phi^c_\bn} \\
\label{eq:exactUt} & &  +  \q{g}\qd{e} \, e^{-i \phi^c_\bn}\; \frac{\sin (\theta^c_{\bn}/2)}{\sqrt{\bn}}\, \ba\daag \, .
\end{eqnarray}
Here, $\theta_n^c \in [0,2\pi)$ is defined by
\begin{equation} \label{eq:etan}
\cos (\theta^c_n/2) = \cos (\theta^{r}_n/2) \; \cos \xi_n\ .
\end{equation}
Introducing~\cite{FNchi}
\begin{equation}\label{eq:defchi}
\chi^c_n = \text{angle}[\,\sin(\theta^r_n/2) - i \cos(\theta^r_n/2) \sin \xi_n \,]\;,
\end{equation}
we define the composite phase as $\phi^c_\bn = \phi_\bn+\chi^c_\bn$.

Comparing Eqs.~(\ref{eq:exactUt}) and (\ref{eq:Utsimp}), we finally get:
\begin{equation}\label{eq:partsUt}
	\bU_{c} =  {\bf Z}(-\phi^c_\bn) \;\, {\bf Y}(\theta^c_\bn) \;\, {\bf Z}(\phi^c_\bn) \, .
\end{equation}
This expression of $\bU_c$ has the same general form as that used in the dispersive case (Section \ref{sec:LargeDetun}). Angles $\theta^c_\bn$, $\phi^c_\bn$ replace $\theta^{r}_\bn$, $\phi^d_\bn$ respectively. We now show that with these adaptations, most of the conclusions of the previous Sections still hold. The reservoir in realistic situations indeed stabilizes the nonclassical states $\q{\psi} \approx e^{-i t_K {\bf H}_K} \, \q{\alpha}$.


\subsection{Effects of ${\bf Y}(\theta^c_\bn)$ and ${\bf Z}(\pm \phi^c_\bn)$}

Let us first consider a reservoir of atoms whose interaction with the cavity would be described by ${\bf Y}(\theta^c_\bn)$. Note that this situation is not physical: the ${\bf Y}(\theta^c_\bn)$ evolution operator is no more than a convenient mathematical factor appearing in the expression of the complete evolution operator $\bU_{c}$.

In analogy with Section \ref{sec:Coherent}, the pointer state $\q{\psi_\infty} = \sum \, \psi_n \q{n}$ corresponding to this fictitious interaction is defined by the recurrence relation:
\begin{equation}
\label{eq:comp:iter}
	\psi_{n+1} \; = \; \frac{\tan (u/2)}{\tan (\theta^c_{n+1}/4)} \, \psi_n\;,
\end{equation}
for $n=0,1,2,...\;$.
Equation \eqref{eq:etan} ensures $\vert \cos(\theta^c_{n+1}/2) \vert < 1$ $\forall n$, therefore $0 < \theta^c_{n+1}/4 < \pi/2$. Moreover $\lim_{n\mapsto +\infty} \theta_n^c = \pi$, such that Eq.~(\ref{eq:comp:iter}) always yields a well-defined finite energy state as soon as $\vert \text{tan}(u/2) \vert < 1$ i.e.~$\vert u \vert < \pi/2$. For large $n$ values, the recurrence~\eqref{eq:comp:iter} is approximated by $\psi_{n+1} =  \text{tan} \tfrac{u}{2} \, \psi_n$ and $\psi_n$ converges exponentially towards $0$ with $\sum_n\; n \, \psi_n^2$ finite.
The energy exchange resulting from the fast commutation of the atomic frequency near the cavity center thus
removes the possibility of trapping states.

Note that even in the absence of the central resonant interaction, with $\theta^{r}_n=0$ in Eq.~(\ref{eq:etan}), relation (\ref{eq:comp:iter}) defines a unique pointer state with finite energy. It is thus in principle possible to simplify our scheme by using only two dispersive phases with opposite detunings.

For $\theta_r$ and $\Omega_r/\delta_0$ small, $2 \tan(\theta^c_n/4) \approx \theta^c_n/2 \approx \theta_c \sqrt{n}/2$ with $\theta_c = \sqrt{\theta_r^2 + (\tfrac{2\Omega_r}{\delta_0})^2}$. The pointer state is thus close to a coherent state $\q{\alpha_\infty}$, as in Section \ref{sec:Coherent}, with the amplitude $\alpha_\infty = 4 \tan(u/2) \, / \, \theta_c$. Convergence arguments similar to those of Section \ref{sec:Coherent} (effective Lindblad master equation) can be given. The convergence rate is now proportional to $\theta_c^2$. We conjecture that this convergence is valid for any $u$ with $0\le u<\pi/2$, $\theta_r \geq 0$, $\Omega_r>0$ and $\delta_0 >0$.

\begin{figure}
		\begin{center}
\includegraphics[width=80mm, trim=0cm 3.8cm 9cm 0.8cm,clip=true]{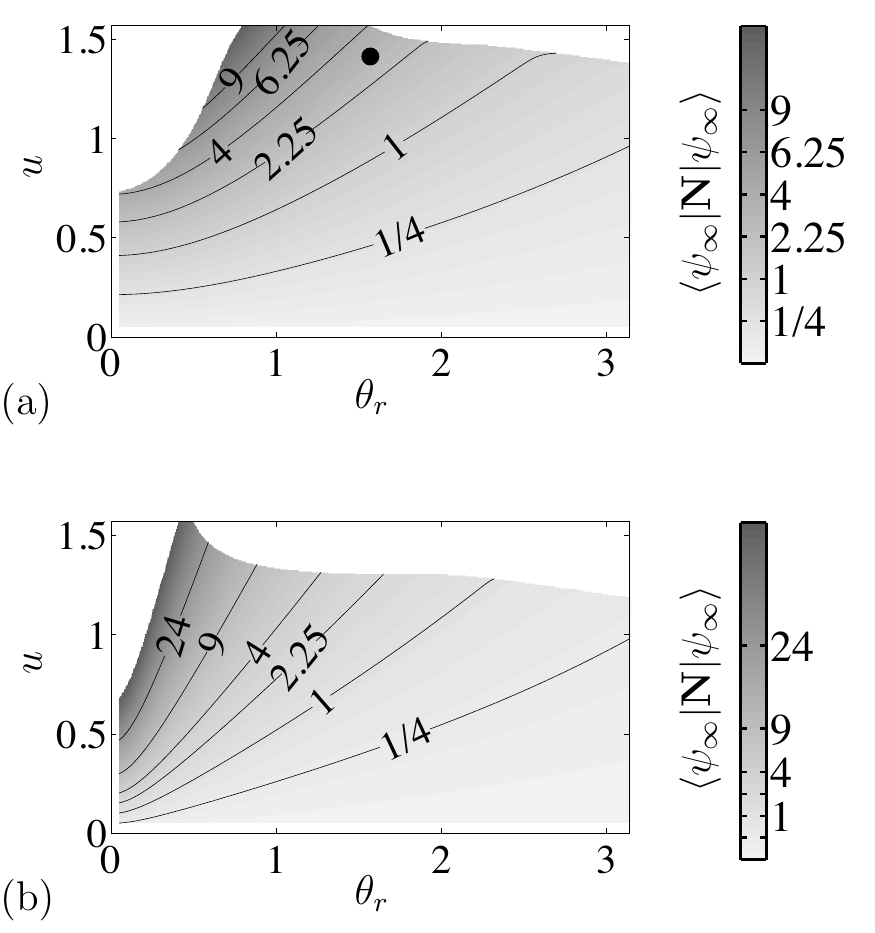}
	\end{center}
\caption{Mean photon number $\qd{\psi_\infty} \bn \q{\psi_\infty}$ of the pointer state $\q{\psi_\infty}$ stabilized by ${\bf Y}(\theta^c_\bn)$, with $\delta_0 = 2.2\, \Omega_r$ (a) and $\delta_0 = 10\, \Omega_r$ (b). Grayscale axis is linear in $\sqrt{\qd{\psi_\infty} \bn \q{\psi_\infty}}$. The 2-component MFSS on Fig.~\ref{fig:Hkerreds} uses $u=0.9\pi/2$ and $\theta_r=\pi/2$ with $\delta_0 = 2.2 \,\Omega_r$, corresponding to the black dot on (a), for which $\qd{\psi_\infty} \bn \q{\psi_\infty} = 2.96$. The shaded zone is delimited such that all corresponding states have at least $99\%$ fidelity $\vert \langle \psi_\infty \q{\alpha_*} \vert^2$ to a coherent state $\q{\alpha_*}$ of same mean photon number ($\vert \alpha_* \vert^2 = \qd{\psi_\infty} \bn \q{\psi_\infty}$).}\label{fig:cohcomp2}
\end{figure}

Figure \ref{fig:cohcomp2} presents numerical estimations of the field pointer state $\q{\psi_\infty}$ stabilized by a hypotetical reservoir using interaction ${\bf Y}(\theta^c_\bn)$. For all represented parameter values, fidelity $\vert \qd{\psi_\infty} \alpha_* \rangle \vert^2$ to a coherent state $\q{\alpha_*}$ of the same mean photon number ($\vert \alpha_* \vert^2 = \qd{\psi_\infty} \bn \q{\psi_\infty}$), is at least $99\%$. That mean photon number is represented by the grayscale. The larger value $\Omega_r/\delta_0=1/2.2$ used on Fig.~\ref{fig:cohcomp2}(a), does not allow to reach as high mean photon numbers as the value $\Omega_r/\delta_0=1/10$ of Fig.~\ref{fig:cohcomp2}(b). Small $\Omega_r/\delta_0$ however are more subject to undesired population of the high-lying Fock states, reminiscent of the trapping states, for large $\theta_r$ and $u$. This explains the smaller domain where fidelity is larger than 99\%. With the particular values used for Fig.~\ref{fig:Hkerreds}(e) (black dot on Fig.~\ref{fig:cohcomp2}(a), with $\Omega_r/\delta_0=1/2.2$, $\theta=\pi/2$ and $u=0.45\pi$), fidelity to a coherent state is almost $99.9\%$ and $\qd{\psi_\infty} \bn \q{\psi_\infty}=2.96$.

We now examine the influence of the ${\bf Z}(\pm \phi^c_\bn)$ operators on the pointer state defined by ${\bf Y}(\theta^c_\bn)$.
A first observation is that it does not modify the photon number populations, since it commutes with $\bn$. Thus, the energy of the field pointer state, for a reservoir with composite interaction, is entirely determined by the parameters in ${\bf Y}(\theta^c_\bn)$, as represented for example on Fig.~\ref{fig:cohcomp2}. Let us define the Hermitian operator $h^c_\bn$ by the recurrence relation:
\begin{equation}\label{eq:hrecurrence}
	h^c_{n+1}-h^c_n = \phi^c_{n+1}\;,
\end{equation}
for $n=0,1,2,...$, with an arbitrary $h^c_0$.  Using Eq.~(\ref{eq:NcommA}) as in Section \ref{sec:LargeDetun} yields
\begin{equation}\label{eq:CV}
\bU_c = e^{-i h_\bn^c} \;\, {\bf Y}(\theta^c_\bn) \;\, e^{i h_\bn^c}\ .
\end{equation}
The pointer states of $\bU_c$ are those of ${\bf Y}(\theta^c_\bn)$ transformed by the unitary $e^{-i h_\bn^c}$. The choice of $h_0^c$ for solving Eq.~(\ref{eq:hrecurrence}) is thus physically irrelevant, as it corresponds in Eq.~(\ref{eq:CV}) to two constant opposite phases that cancel out.
The operator $h^c_\bn$ here plays exactly the role of $h^d_\bn$ in Section \ref{sec:LargeDetun}. The only difference is that, as $\phi^c_n$ is nonlinear, $h^c_n$ is defined with the discrete integral \eqref{eq:hrecurrence}. If $\phi^c_n$ is nearly linear in $n$ over the relevant photon numbers [dominant photon numbers in the pointer state $\q{\psi_{\infty}}$ associated to ${\bf Y}(\theta^c_\bn)$], then $h^c_n$ is nearly quadratic and the situation of Section \ref{sec:ArbitDetun} is recovered. The reservoir stabilizes nonclassical pointer states $\q{\psi^c_\infty} = e^{-i h_\bn^c} \, \q{\psi_\infty} \approx e^{-i t_K {\bf H}_K} \q{\alpha}$ with $t_K$ depending on the parameters governing $\phi^c_n$.

\subsection{Choice of the reservoir operating point}

We now use this detailed description of the reservoir to justify the choice of operating parameters leading to the generation of the two component MFSS presented in figure \ref{fig:Hkerreds}(e): $u=0.45\,\pi$, $\theta_r=\pi/2$, $v=70$~m/s, $\delta=2.2\,\Omega_0$. Note that the state in Fig.~\ref{fig:Hkerreds}(e), with $\approx 2.7$ photons on the average has been computed with a finite cavity lifetime $T_c=65$~ms. The same computation in an ideal cavity leads to an average photon number equal to 2.96, see Fig.~\ref{fig:cohcomp2}. The \emph{two-component} MFSS corresponds to the largest effect of the dispersive interaction, and hence to the most demanding experimental conditions.

The chosen parameters are the result of a tradeoff between contradictory requirements. First, the composite phase shift $\phi^c_n$ must be nearly linear in $n$ over the useful photon number range, with a slope of $\pi$ per photon. Second, the time of convergence towards the steady state needs to be much shorter than the decoherence time ($T_c/5.6$) of the target state due to unavoidable cavity relaxation. Linearity of $\phi^c_n$ improves with larger $\delta_0/\Omega_r$. The $\pi$ phase shift per photon condition then requires very long atom-cavity interaction time, in clear contradiction with the second requirement.

\begin{figure}
		\begin{center}
\includegraphics[width=80mm, trim=0cm 4.2cm 9cm 0.5cm,clip=true]{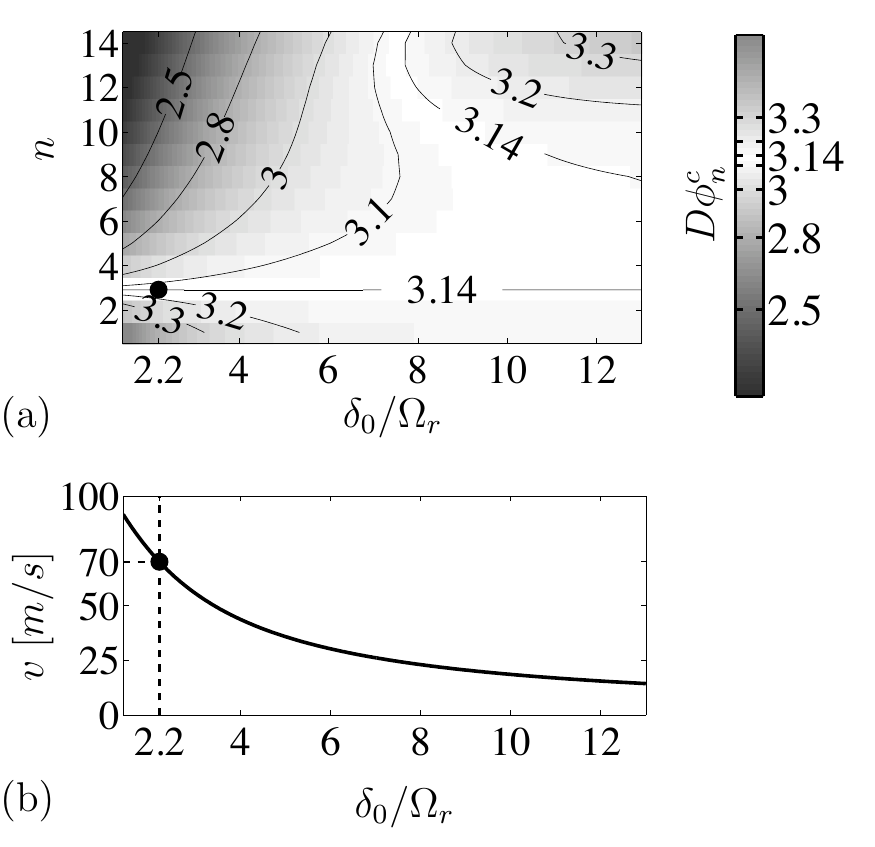}
	\end{center}
\caption{(a) Kerr-effect-inducing $D \phi^c_n = \phi^c_{n+1}-\phi^c_n$ as a function of photon number $n$. Since $\phi^c_0$ is undefined, we start with $D \phi^c_1$. We set $\theta_r=\pi/2$. (b) Corresponding velocities $v$: for each $\delta_0/\Omega_r$, we adjust $v$ to have $D\phi^c_n=\pi$ at $n=2.96$. That value is chosen to cover the parameter values $v=70$~m/s, $\delta_0/\Omega_r=1/2.2$ (black dots) used for the 2-component MFSS in Fig.~\ref{fig:Hkerreds}. An ideal $h^c_n$, proportional to ${\bf H}_K$, requires $D\phi^c_n$ constant in $n$. Small $\delta_0/\Omega_r$ values are disadvantageous for this criterion, but allow higher velocity and hence more frequent reservoir atoms for a same mean $D\phi^c$.}\label{fig:DH2}
\end{figure}

The tradeoff is further examined on Fig.~\ref{fig:DH2}. Figure \ref{fig:DH2}(a) evaluates the linearity of $\phi^c_n$ by showing $D \phi^c_n = \phi^c_{n+1}-\phi^c_n = h^c_{n+1}+h^c_{n-1} - 2 h^c_n$ for different parameter values. Once again $\theta_r$ has little influence and we set it to $\pi/2$. For each value of $\Omega_r/\delta_0$, we adjust $v$ to have $D \phi^c_n = \pi$ for the same mean photon number $n=2.96$ (by interpolation). That value is chosen to cover the parameter values of Fig.~\ref{fig:Hkerreds}(e), represented by a black dot on Figs.~\ref{fig:cohcomp2} and \ref{fig:DH2}. As expected, $D \phi^c_n$ is quite constant for moderate photon numbers in the dispersive region $\delta_0/\Omega_r\gg 1$. This corresponds however to unrealistically small atomic velocities, represented on Fig.~\ref{fig:DH2}(b). In the region of low $\delta_0/\Omega_r$ values, a $D \phi^c_n \approx \pi$ at $n=2.96$ can be reached with larger $v$, but $D \phi^c_n$ varies more rapidly with $n$. This variation is nevertheless sufficiently weak in the range $2\le n\le 5$ for $\delta_0/\Omega_r\approx 2.2$, corresponding to the $v=70$~m/s that is used for Fig.~\ref{fig:Hkerreds}(e).

\begin{figure}
		\begin{center}
\includegraphics[width=80mm, trim=0cm 0.5cm 9.5cm 0.5cm,clip=true]{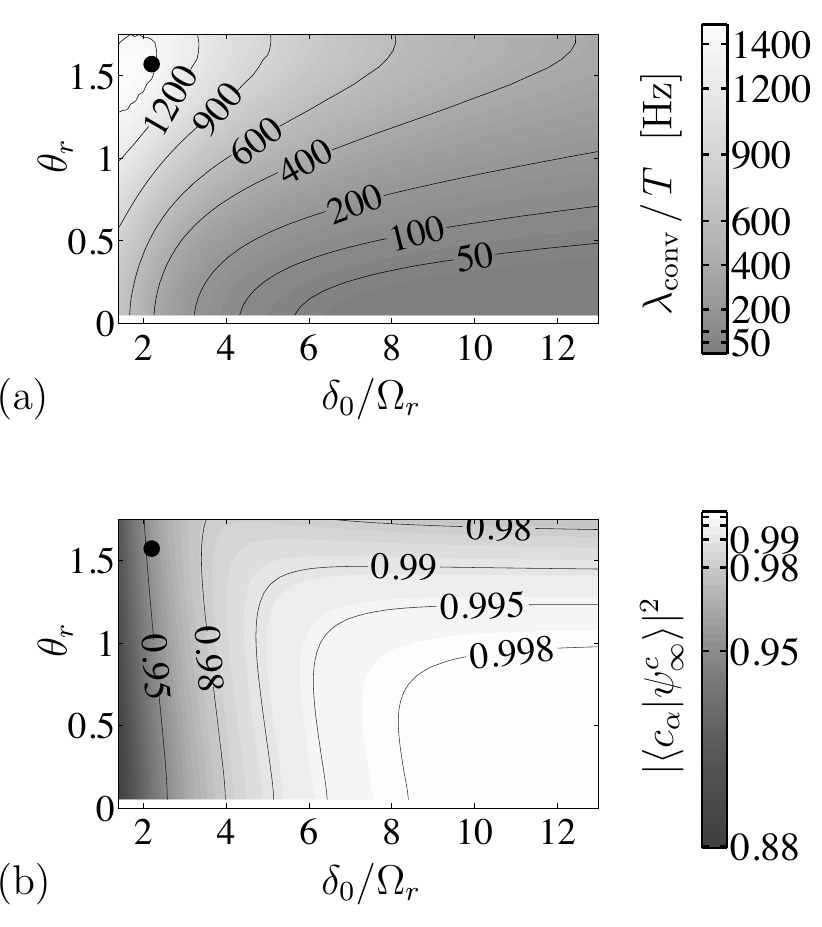}	
	\end{center}
\caption{(a): Convergence rate $\lambda_{\text{conv}} / T$ giving the slope, in time units ($\text{s}^{-1}$), of the convergence towards the reservoir pointer state $\q{\psi_\infty^c}$, according to $\;\log \vert \log \qd{\psi_\infty^c} \rho_k \q{\psi_\infty^c}\vert = \log \vert \log \qd{\psi_\infty^c} \rho_0 \q{\psi_\infty^c}\vert - \lambda_{\text{conv}}\, k\;$ (see Fig.~\ref{fig:CohEvMap}). For each $\theta_r$ and $\Omega_r/\delta_0$, we adjust $v$ as in Fig.~\ref{fig:DH2} to keep $D \phi^c_n \approx \pi$, and $u$ to keep $\qd{\psi_\infty^c} \bn \q{\psi_\infty^c} = 2.96$; this reference comes from the values $\theta_r=\pi/2$, $u=0.45 \pi$, $\Omega_r/\delta_0=2.2$, $v=70$~m/s (black dot) used for Fig.~\ref{fig:Hkerreds}(e). Time $T=3w/v$ between consecutive atoms changes as we adjust $v$. (b): Fidelity of the same $\q{\psi_\infty^c}$ to a 2-component MFSS $\q{c'_{\alpha_\infty}} = (\q{\alpha_\infty}+i e^{i\beta}\, \q{\minou\alpha_\infty})/\sqrt{2}$, where we tune $\alpha_\infty$ and $0 \leq \beta < \pi$ to optimize fidelity. It turns out that $\vert \beta \vert < 0.005$ for most parameter values, while $\vert \alpha_\infty \vert^2$ decreases as fidelity decreases, below 2.7 for the lowest values of $\Omega_r/\delta_0$. The black dot marks the case of the 2-component MFSS in Fig.~\ref{fig:Hkerreds}. For $\theta_r$ values larger than those represented, no $u$ value stabilizes a mean photon number $2.96$, see also Fig.~\ref{fig:cohcomp2}. The two plots together illustrate a tradeoff between fidelity in absence of decoherence and convergence speed.}\label{fig:Fidcomp}
\end{figure}

Let us now examine the overall reservoir fidelity and the convergence rate $\lambda_{conv}$ from the vacuum towards the target state, as defined in Fig.~\ref{fig:CohEvMap}. We choose as free reservoir parameters $\theta_r$ and $\delta_0/\Omega_r$.
This choice sets the value of $v$ [see Fig. \ref{fig:DH2}(b)]. Then $u$ is adjusted so that the target mean photon number is $2.96$ (see Fig.~\ref{fig:cohcomp2}(a)). Figure \ref{fig:Fidcomp}(a) shows the ratio $\lambda_{conv}/T$, where $T$ is the total interaction time of each atom with the cavity. This ratio is the real convergence rate in $\text{s}^{-1}$ units. For Fig.~\ref{fig:Hkerreds}(e), since the expected target state decoherence time is of order $65/5.6\approx 10$~ms, we choose a parameter set $\delta_0/\Omega_r=2.2$ and $\theta_r=\pi/2$, corresponding to a $1400$~s$^{-1}$ convergence rate (black dot on Fig.~\ref{fig:Fidcomp}(a)). The fidelity with respect to an ideal MFSS with the same energy is shown on Fig.\ref{fig:Fidcomp}(b). Our choice of parameters does not correspond to a maximum fidelity due to the variation of $D\phi^c_n$ in the useful $n$ value range (see Fig.~\ref{fig:DH2}(a)). However, we get an excellent $95\%$ fidelity.

\begin{figure}
		\begin{center}
\includegraphics[width=80mm]{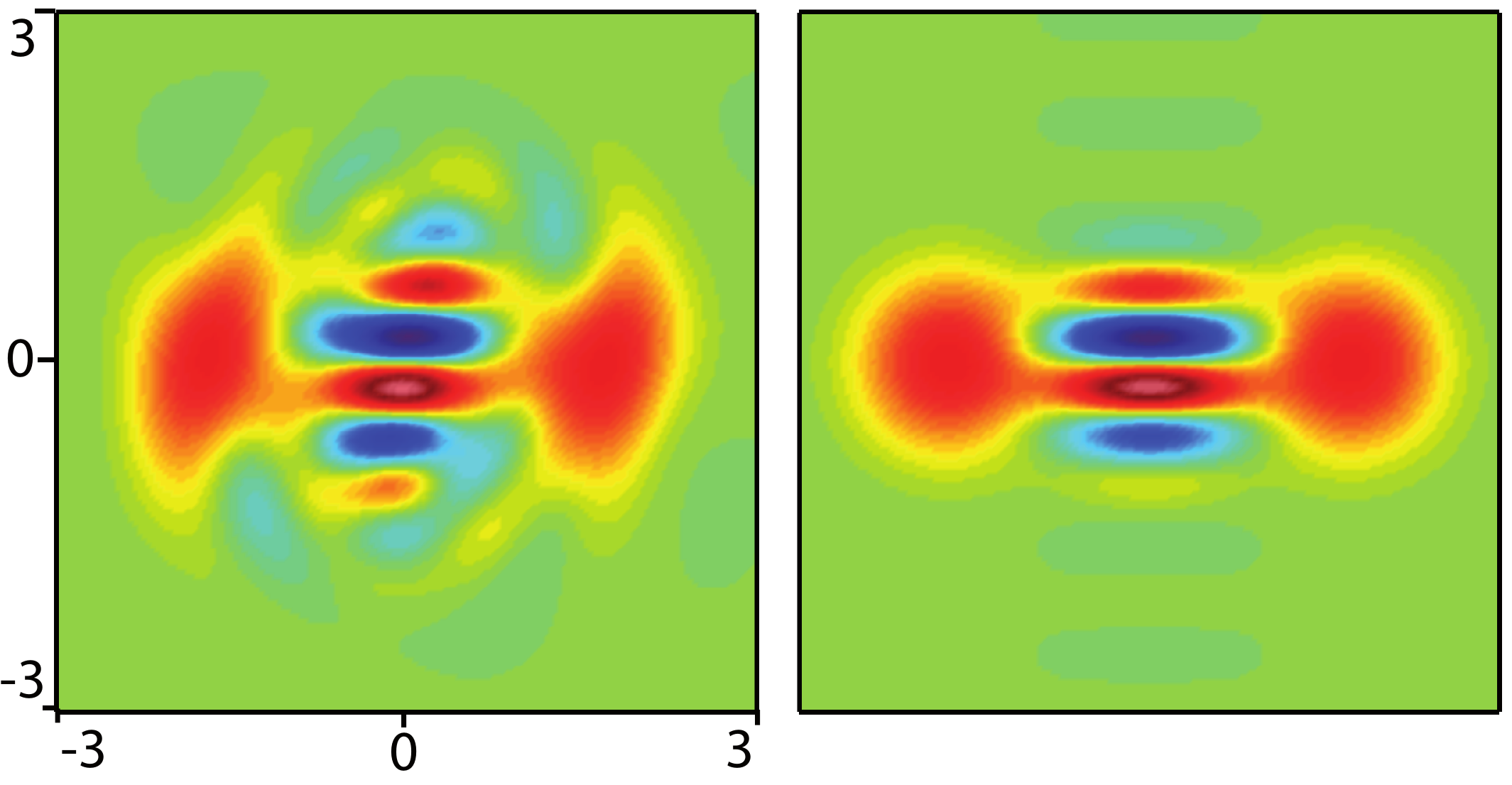}
	\end{center}
\caption{Wigner function illustrating a stabilized 2-component MFSS (colorbar as in Fig.~\ref{fig:Hkerreds}). Left: $\q{\psi_\infty^c}\qd{\psi_\infty^c}$, pointer state of our reservoir with composite interaction. Parameter values are $\theta_r=\pi/2$, $u=0.45\pi$, $\delta_0=2.2\Omega_0$, $v=70$~m/s, i.e.~those used for figure \ref{fig:Hkerreds}(e), except $T_c$ set to infinity here. Right: target state $\q{c_\alpha}\qd{c_\alpha}$.}\label{fig:new}
\end{figure}

Figures \ref{fig:new} left and right respectively present the Wigner functions of the steady state MFSS obtained with this parameter choice, and of a theoretical superposition of two coherent states with opposite phases and same total energy. The slight distortions of the quasi coherent components in the pointer state MFSS are due to the non-linearity of the phase shift $\phi^c_n$.

\section{Decoherence and experimental imperfections}\label{sec:decoherence}
The choice of operating parameters performed above has been based on a rough estimate of the action of decoherence.
We now show how the reservoir allows us to stabilize MFSS with a high fidelity, in presence of cavity relaxation due to a zero-temperature environment (Section \ref{ssec:photonloss}). In Section \ref{ssec:ximp}, we study the robustness of the scheme against other experimental imperfections by numerical simulations.

\subsection{Cavity relaxation}\label{ssec:photonloss}

First, we consider the field evolution with a simplified model. It is obtained from equation \eqref{eq:Zdyns} for a coherent evolution, sandwiched between two dispersive transformations (Eq.~\eqref{eq:DispIntZ}):
\begin{eqnarray}
  	\alpha_{k+1} &=& (1-\theta_r^2 / 8) \alpha_k + u \theta_r / 4 \notag\\
    \rho'^h_k&=&\q{\alpha_k}\qd{\alpha_k}\notag\\
    \rho'_k&=&e^{-i\pi/2~\bn^2}\rho'^h_k e^{i\pi/2~\bn^2}\;.\label{eq:simplified_model}
\end{eqnarray}
In a Monte-Carlo approach, the evolution of the field density matrix due to cavity relaxation can be represented as a succession of quantum jumps described by the annihilation operator $\ba$, occurring at random times and interrupting a non-unitary deterministic evolution of the field state \cite{Molmer-Dalibard-JOptSoc_93}.

The action of $\ba$ on an MFSS $\q{c_\alpha}$ writes:
\begin{equation*}
 \ba\q{c_\alpha}\propto \q{c_{-\alpha}} \, .
 \end{equation*}
Since the loss of photons in the environment cannot be detected, an initial $\q{c_\alpha}$ state rapidly evolves in the absence of the reservoir into a statistical mixture of $\q{c_\alpha}$ and $\q{c_{\minou\alpha}}$, i.e. into a mixture of $\q{\alpha}$ and $\q{\minou\alpha}$.
When the reservoir is present, it drives $\q{c_{\minou\alpha}}$ back to $\q{c_{\alpha}}$ after each jump. If the reservoir-induced convergence time is much shorter than the average interval between two jumps, then the field is mostly close to $\q{c_{\alpha}}$.

This simple description suggests to seek a solution for the steady state with decoherence under the form $\rho'^h_\infty=\int{ \mu(z) \q{z}\qd{z}~ dz}$.
This is a statistical mixture of coherent states $\q{z}$ with real amplitudes $z$ weighted by $\mu(z)$.

In the absence of cavity relaxation, the evolution of $\rho'^h_k$ in the simplified model [Eq.~\eqref{eq:simplified_model}] can be viewed as a discretization of the Lindblad master equation:
\begin{equation}\label{eq:cont-approx}
	\tfrac{d}{dt}\rho'^h = [\beta \ba\daag-\beta\daag \ba,\, \rho'^h] - \tfrac{\kappa}{2} (\bn \rho'^h + \rho'^h \bn - 2 \ba \rho'^h \ba\daag) \, ,
\end{equation}
with $\beta \, dt = u \theta_r / 4$ and $\kappa\, dt = \theta_r^2 / 4$. Eq.~\eqref{eq:cont-approx} describes the evolution of the field mode coupled with a classical source with an amplitude $\beta$ and damped at a rate $\kappa$. At long times, $\rho'$ converges towards a coherent state $\q{\alpha_{\infty}}$ with $\alpha_\infty = 2 \beta / \kappa$, see e.g.~\cite{HarocheBook}. Note that in Eq.\eqref{eq:cont-approx}, the damping rate $\kappa$ is induced by \emph{the atomic reservoir} and not by cavity relaxation.
Since $\rho'^h$ follows \eqref{eq:cont-approx}, $\rho'$ follows
\begin{multline*}
	\tfrac{d}{dt}\rho' = \beta [ \ba\daag e^{-i\pi\bn} -e^{i\pi\bn}\ba,\, \rho'] \\- \tfrac{\kappa}{2} (\bn \rho' + \rho' \bn - 2 e^{i\pi\bn}\ba \rho' \ba\daag e^{-i\pi\bn}) \ ,
\end{multline*}
where we can assume, up to a change of phase, that $\beta$ is real and positive.

We now add to this simple model a thermal environment at zero temperature that induces decoherence of the field with the cavity lifetime $T_c = 1/\kappa_c$. This adds the usual Lindblad terms to the second member of the previous equation and $\rho'$ now obeys:
\begin{multline}
	\tfrac{d}{dt}\rho' = \beta [ \ba\daag e^{-i\pi\bn} -e^{i\pi\bn}\ba,\, \rho'] \\- \tfrac{\kappa}{2} (\bn \rho' + \rho' \bn - 2 e^{i\pi\bn}\ba \rho' \ba\daag e^{-i\pi\bn}) \\
- \tfrac{\kappa_c}{2} (\bn \rho' + \rho' \bn - 2 \ba \rho' \ba\daag ) \,. \label{eq:rhoDecoherence}
\end{multline}
In the Kerr representation, $\rhoK$ then evolves according to:
\begin{multline}
	\tfrac{d}{dt}\rhoK = \beta [ \ba\daag-\ba,\, \rhoK] \\
  - \tfrac{\kappa+\kappa_c}{2} (\bn \rhoK + \rhoK \bn - 2 \ba \rhoK \ba\daag ) \\
  - \kappa_c (\ba \rhoK \ba\daag - e^{i\pi\bn}\ba \rhoK \ba\daag e^{-i\pi\bn} ) \, .\label{eq:rhoKDecoherence}
\end{multline}
Without the terms in the third line of Eq.~\eqref{eq:rhoKDecoherence}, we would get Eq.~\eqref{eq:cont-approx} with $\kappa$ replaced by $\kappa+\kappa_c$. This would yield a coherent steady state of amplitude $\alpha^c_\infty = \alpha_\infty / (1+\eta)$ with $\eta = (4 T)/(\theta^2_r\, T_c)$. The whole equation \eqref{eq:rhoKDecoherence} leaves invariant the set of mixtures of coherent states with real amplitudes in $[-\alpha^c_\infty,\,\alpha^c_\infty]$. We therefore search for the stationary solution under the form:
\begin{equation}\label{eq:rhoKinfty}
\rhoK_\infty = \int_{-\alpha^c_\infty}^{\alpha^c_\infty} \mu(z) \q{z}\qd{z}~ dz \, .
\end{equation}
As shown in the Appendix~\ref{sec:dampeq}, this yields a solution:
\begin{equation}\label{eq:mu}
\mu(z) = \mu_0\, \frac{\left(((\alpha^c_\infty) ^2 - z^2)^{(\alpha^c_\infty)^2}~e^{ z^2}\right)^{r_c}}{\alpha^c_\infty  - z}\ ,
\end{equation}
with $r_c=2\kappa_c/(\kappa+\kappa_c)$. The normalization factor $\mu_0>0$ ensures that $ \int_{-\alpha^c_\infty}^{\alpha^c_\infty} \mu(z) dz \, =1$. In any case, $\mu(-\alpha^c_\infty ) =0$. For small $\kappa_c$, we have $\lim_{z\mapsto \alpha^c_\infty}\mu(z)=+\infty$ and $\rhoK_\infty$ is close to the coherent state $\q{\alpha^c_\infty}$. For large $\kappa_c$, $\alpha^c_\infty$ tends to zero and thus the field steady-state becomes close to the vacuum.

\begin{figure}
	\begin{center}
\includegraphics[width=80mm]{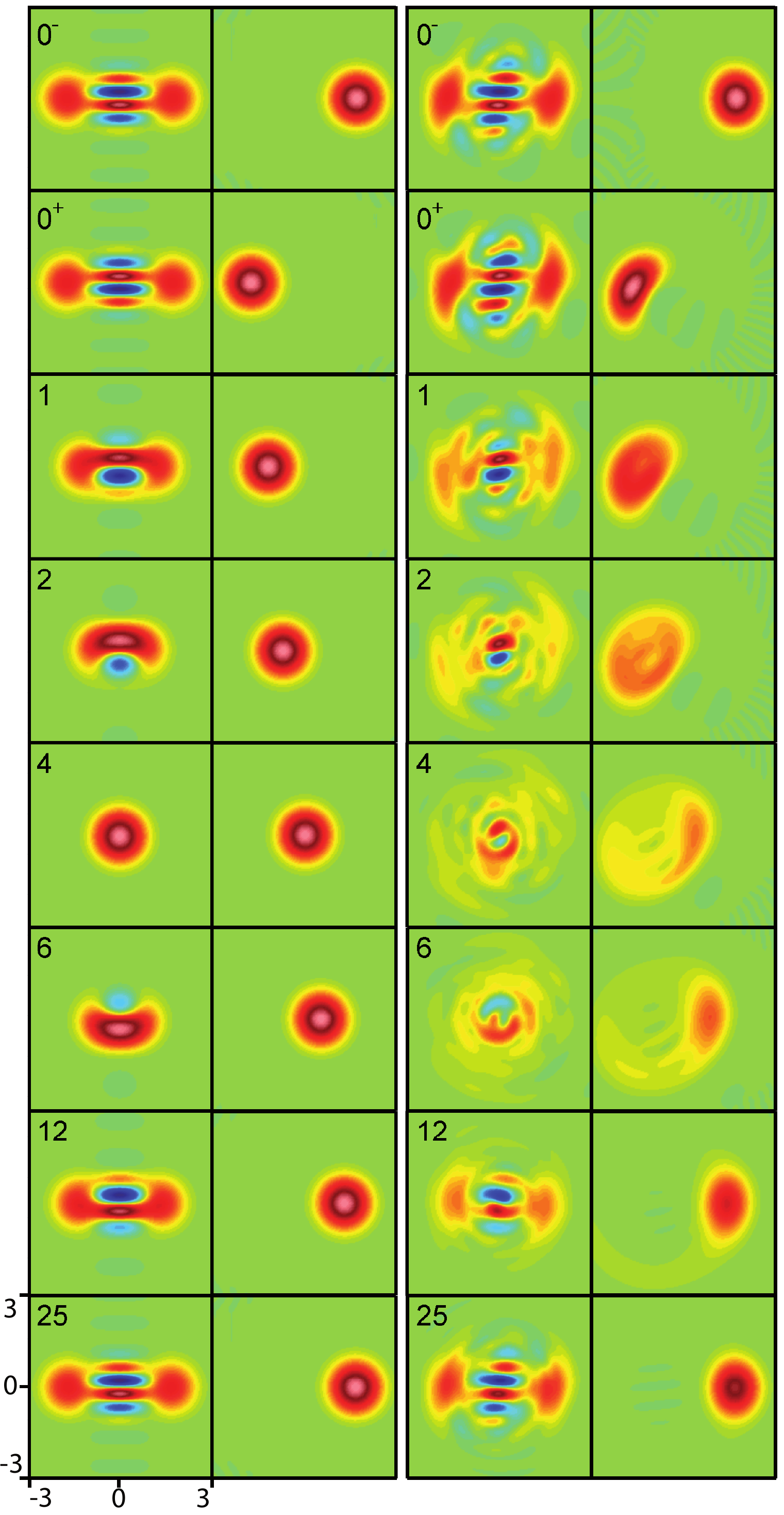}
\end{center}
	\caption{Evolution of the cavity field coupled to a reservoir stabilizing a 2-component MFSS, immediately after a relaxation-induced photon loss. Parameter values are those used for Fig.~\ref{fig:new}. The frames are labelled by the number of atomic interactions. A photon loss out of the reservoir pointer state occurs between frames labelled $0^-$ and $0^+$.
Left two columns: simplified model, described by Eq.~\eqref{eq:simplified_model}. We show the Wigner functions of both the cavity state, $\rho'$, on column 1 and of $\rho'^h$ on column 2. Right two columns: same plots for the actual reservoir characterized by $\bU_c$ ($\rho$ on column 3 and $\rho^h$ on column 4). }\label{fig:filmcatkerr}
\end{figure}

We now compare this simplified model to the actual reservoir in the presence of relaxation.
Figure \ref{fig:filmcatkerr} illustrates the reservoir-induced convergence after a quantum jump.
The leftmost column shows the Wigner function of $\rho'$ during this recovery process for the simplified model \eqref{eq:simplified_model}. We start as $\q{c_{\alpha}}$ (upper left frame). Immediately after a jump (second frame in the leftmost column), the state is $\q{c_{\minou\alpha}}$. Successive snapshots of the recovery procedure are presented in the next frames. We neglect here the action of cavity relaxation during this recovery process. Note that after $\approx 4$ reservoir atoms, the state is the vacuum, from which $\q{c_{\alpha}}$ is gradually recovered.

The second column depicts the evolution of $\rho'^h$. In this representation, the initial state is the coherent state $\q{\alpha}$ (first frame). It jumps to $\q{\minou\alpha}$ (second frame), and then gradually evolves back towards $\q{\alpha}$ according to Eq.~\eqref{eq:Zdyns}, staying coherent at all time.

On the third column, we show the Wigner functions of the actual cavity state $\rho$ induced by our reservoir, whose dynamics is governed by the Kraus map associated to $\bU_c$.
The last column shows the evolution of $\rho^h = e^{i h_\bn^c} \, \rho \, e^{-i h_\bn^c}$.
We observe that $\rho^h$ and $\rho'^h$ follow qualitatively the same path. The main difference is a notable distortion of $\rho^h$  when the field amplitude is near zero.

\begin{figure}
		\begin{center}
\setlength{\unitlength}{1mm}
\begin{picture}(80,126)
\put(0,0){\includegraphics[width=75mm]{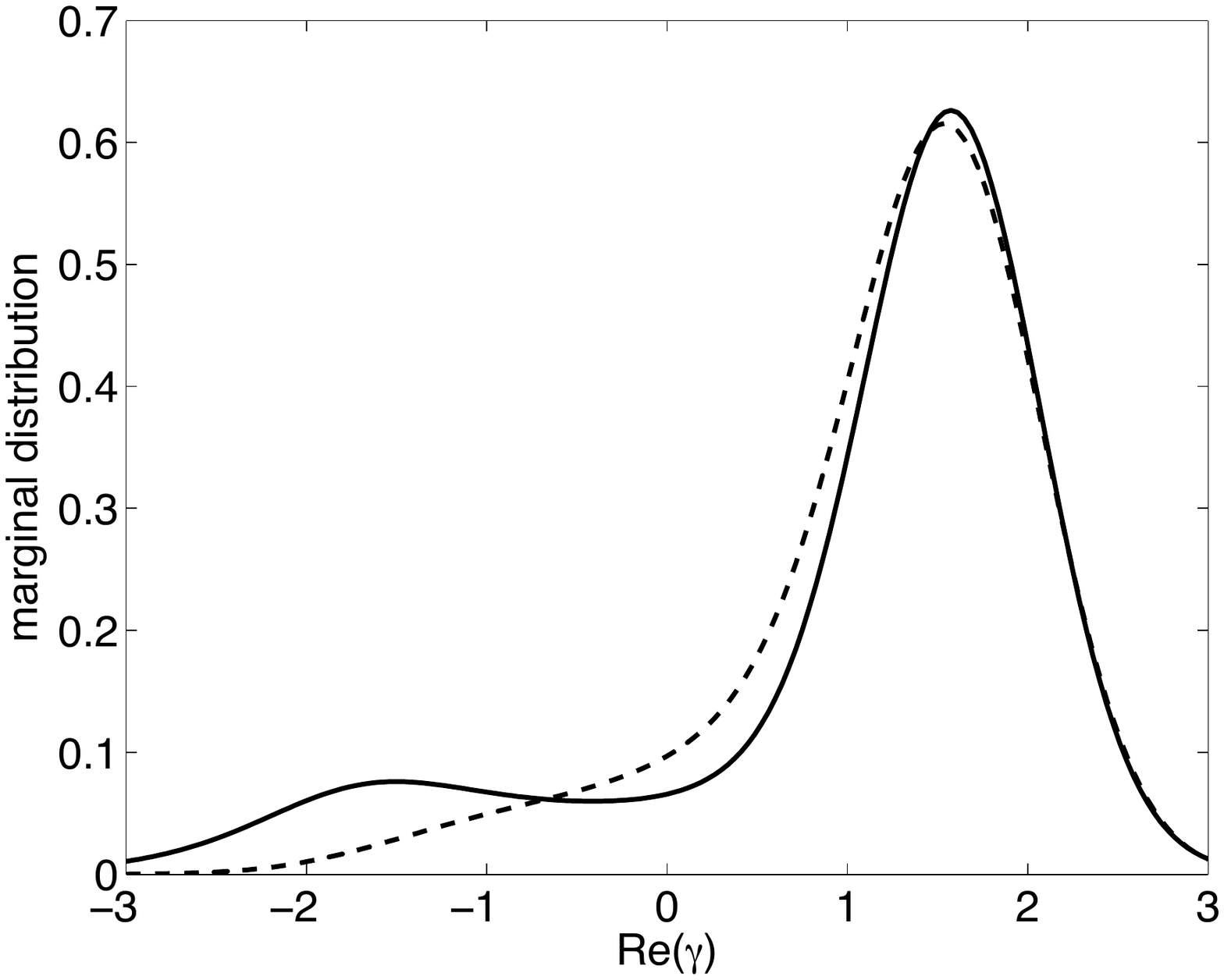}}
\put(0,63){\includegraphics[width=75mm]{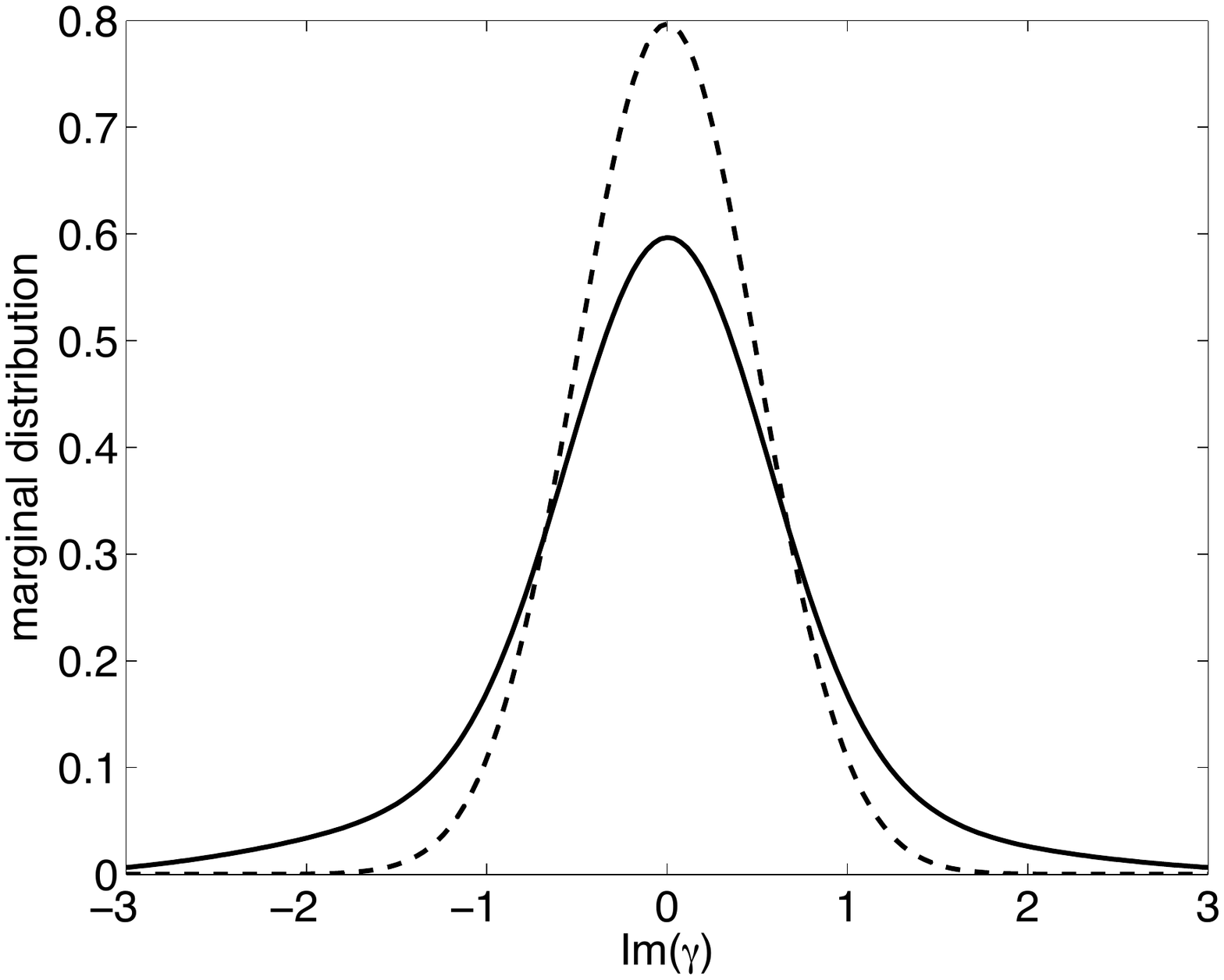}}
\end{picture}
	\end{center}
\caption{Steady state of the cavity field coupled to the atomic reservoir and to a relaxation-inducing environment with $T_c=$65~ms.
Top [resp: bottom]: marginal distribution of the Wigner function along the imaginary [resp: real] quadrature for the simplified model (dashed line) and for our reservoir (solid line). These states correspond to the quantum Monte Carlo average of the sequence presented in Fig.~\ref{fig:filmcatkerr}. }\label{fig:previousAvd}
\end{figure}

In figure \ref{fig:previousAvd}, we plot the two marginal distributions of the Wigner functions for $\rho'^h_\infty$ and $\rho^h_\infty$ along the real and complex quadratures. The reservoir steady states $\rho'^h_\infty$ and $\rho^h_\infty$ approximately correspond to the quantum Monte Carlo average of the trajectories depicted in Fig.~\ref{fig:filmcatkerr}. Figure \ref{fig:previousAvd} features dominant peaks which suggest that the field is mostly close to the target. The distortions with respect to a coherent state visible on the fourth column of Fig.~\ref{fig:filmcatkerr}, lead to a plateau or bump on the marginal distributions of $\rho^h_\infty$. We nevertheless observe that our simplified model [Eq.\eqref{eq:rhoDecoherence}] captures the main features of the influence of decoherence.

\subsection{Experimental uncertainties}\label{ssec:ximp}

We have performed extensive numerical simulations to assess the robustness of the reservoir  versus uncertainties in the experimental settings. The evolution operators during the interaction are computed exactly from the Hamiltonian ${\bf H}_{JC}$ [Eq. (\ref{eq:JCm})], using the quantum optics package for MATLAB~\cite{Quopackage}. Hilbert space is truncated to the 60 first Fock states. We take into account separately the atom-cavity coupling ruled by ${\bf H}_{JC}$ and the relaxation of the cavity mode (modeled in the standard Lindblad form). This simplifying approach holds since $T\ll T_c$.

We use as a reference the generation of a two-component MFSS containing 2.7 photons on the average (thermal environment with $T_c=65$~ms and a mean number $n_t=0.05$ of blackbody photons per mode, $\delta=2.2\,\Omega_0$, $v=70$~m/s, $u=0,45\pi$, $t_r=5\ \mu$s, see Fig.~\ref{fig:Hkerreds}(e)). We take into account the randomness of the Rydberg state preparation~\cite{PRancestor}. In each atomic sample, we excite a random number of atoms obeying a Poisson law with an average $p_{at}$. We take a low value $p_{at}=0.3$ such that, in a first approximation, we only get samples containing zero or one atom (we examine later in this Section the influence of samples containing two atoms).  Note that these are the conditions used for figure \ref{fig:Hkerreds}(e). In each case, we compute, for a slight change in the experimental settings, the variation of the fidelity of the pointer state w.r.t.~an ideal optimized two-component MFSS. For the reference set of parameters, the steady-state fidelity is 70\%.

For velocities in the $66\le v\le 74$~m/s interval, the fidelity is only slightly altered, varying from 65\% to 70\%. It is thus insensitive to a velocity dispersion in the 10\% range, well below the values achieved in the experiment.

The fidelity is also quite insensitive to a slight mismatch in the values of the detuning for the two dispersive interactions. Assuming that $\delta$ takes the value $a_1\times 2.2\,\Omega_0$ in the first dispersive period and $-a_2\times 2.2\,\Omega_0$ in the second, the fidelity drops by at most 10\% when $a_1$ and $a_2$ vary by up to 10\%. The latter cover far more than the actual uncertainty on the atomic frequency.

We have slightly offset the timing of the resonant interaction, shifting it in time by $\delta t$ and keeping $t_r$ constant. A shift of up to 1~$\mu$s (well above the 10~ns timing accuracy) has no effect on the fidelity at the 1\% level. The fidelity is also quite insensitive to a finite rise time for the voltage controlling the atomic Stark effect in the cavity, and hence to a finite commutation time for the detuning $\delta$. Using an exponential relaxation model, and setting $t_r$ to maintain a constant $\theta_r$ value, we find that the fidelity is unchanged for commutation times up to 200~ns, in the range of accessible values.

We have also studied the sensitivity to the atomic samples containing two atoms at the same time. We decide randomly for each sample the actual number of atoms, $N_a$, according to a Poisson distribution with the average value $p_{at}$, truncated above $N_a=2$. For two-atom samples, we integrate the exact equations of motion, assuming an identical coupling of both atoms to the mode. This condition is realized in the experiment, since the maximum separation between the atoms in a sample is, below 1~mm  in $C$, much smaller than the wavelength -- 6~mm -- or than the mode waist $w$.

We observe that the two-atom events do have an impact on the fidelity. For $p_{at}=0.3$, the energy of the prepared cat decreases down to 2.4 photons on the average and the fidelity is reduced to 66\%. For larger $p_{at}$ values, the decrease is more important and the fidelity reduces to 34\% for $p_{at}=0.5$ (for larger $p_{at}$, the simulation should also include 3-atom samples).

If we consider an unrealistic reservoir involving samples containing always two atoms, we get as steady-state a large two-component cat, with 4.8 photons on the average and a fidelity of 65\%. In the real situation, this two-atom engineered reservoir interferes destructively with the operation of the one-atom samples, leading to reduced average energy and fidelity.

When we reduce $p_{at}$ below 0.3, the fidelity and the energy also decrease, since the reservoir is then less efficient to counteract decoherence. For $p_{at}=0.2$,  we get a 1.9 photons state with a  fidelity of only 54\%. Optimizing the average number of atoms per sample is thus important to achieve an efficient engineered reservoir.

Note finally that the phase of the MFSS coherent components is determined by the phase of the atomic state superposition when the resonant interaction period begins. Since the atom is detuned from the mode during the dispersive interactions, this phase rotates at frequency $\pm\delta_0$ during the time interval $-T/2  \le t\le -t_r/2$. The timing of Stark shifts, that determines the atom-field interactions, should thus define $(T-t_r)/2$ with an uncertainty much smaller than $1/\delta_0$ to avoid spurious rotations of this phase. With detunings in the 100~kHz to few MHz range, this timing accuracy is easily achieved.

\section{A reservoir for two-mode ESMS}\label{sec:2modes}

Our reservoir engineering strategy can be adapted to protect entangled state superpositions of two cavity modes, which violate a Bell inequality. Preparation of entangled states of two cavity modes, without protection, were considered in \cite{Milman-al-EurPhysJD_2005,Rauschenbeutel-al-PRA_2001}. An approximate reservoir for entangling large atomic ensembles is proposed and realized in~\cite{Krauter2011}. In ion traps, reservoir engineering has recently been used to stabilize a Bell state and a GHZ state of four qubits \cite{Barreiro-et-al:Nature_2011}.

We present here a scheme in which the two modes belong to the same cavity (two TEM modes of orthogonal polarization, whose degeneracy is lifted by an appropriate mirror shape). Extension to two separate cavities would require atoms going back and forth between them, a feat not easily achieved in the present context of the ENS experiments.

\subsection{Model and target}

We consider two modes $a$ and $b$ of the cavity of respective frequencies $\omega_a < \omega_b$. We note $\bb$ [resp.~$\ba$] the photon annihilation operator for mode $b$ [resp.~mode $a$] and $\bn_b=\bb^\dag \bb$ [resp: $\bn_a=\ba^\dag \ba$] the associated photon number operators. A separable joint state of the two modes is written $\q{\psi_a,\psi_b}$.
The atomic qubit (transition frequency $\omega_0\approx\omega_a,\omega_b$) interacts with the modes according to the Jaynes-Cummings Hamiltonian, which writes, in a frame rotating at the frequency $\omega_m = (\omega_a+\omega_b)/2$:
\begin{eqnarray}
  \label{eq:H2modes}
    \overline{\bf H}_{JC} &=& \Delta \, (\bn_b - \bn_a) + \frac{\delta(t)}{2}(\q{e}\qd{e}-\q{g}\qd{g})\\
    && + i\frac{\Omega(s)}{2} (\q{g}\qd{e}(\ba^\dag+\bb^\dag)-\q{e}\qd{g}(\ba+\bb))\;,\notag
\end{eqnarray}
where $\Delta=(\omega_b-\omega_a)/2 >0$ and $\delta(t) = \omega_0(t) - \omega_m$. Here again, $\delta(t)$ can be adjusted by controlling $\omega_0$ through the Stark effect. We assume that the couping  $\Omega(s)$ is the same with both modes, a restriction that could be easily relaxed.

We note $\bbU$ the unitary evolution operator associated to $\overline{\bf H}_{JC}$ (the overline here denotes two-mode operators), solution of the Schr\"odinger equation:
\begin{equation}
\label{eq:Uev2}
   \frac{d}{dt}\bbU(t)=-i \overline{\bf H}_{JC}(t)\, \bbU(t) \quad \text{with }\; \bbU(t_0) = \bid \; .
\end{equation}
We note  $\bbU_q$ the two-mode evolution operator corresponding to the parameter set $q$, and  (${\bf M}_g^{\bbU_q},{\bf M}_e^{\bbU_q}$) the associated Kraus operators. Approximate analytical expressions of $\bbU_q$ for the relevant parameter sets are given in Appendix~\ref{sec:propags}. Operators $\bbZ$ and ${\bf \overline Y}$ generalizing for each mode the ones introduced in the previous Sections are also defined in the Appendix.

Let us consider first the successive resonant interaction of the atoms, intially prepared in $\q{u_{at}}=\cos(u/2)\q{g}+\sin(u/2)\q{e}$, with the modes $b$ and $a$.
The corresponding propagator is $\bbU_r={\bf \overline Y}(\theta^r_{\bn_a}){\bf \overline Y}(\theta^r_{\bn_b})$. The associated Kraus map $\left({\bf M}_g^{\bbU_r},{\bf M}_e^{\bbU_r}\right)$ stabilizes a tensor product of two coherent states $\q{\minou\alpha,\alpha}$, where $\alpha=2u/\theta_r$ for small enough $u$ and $\theta_r$.

Under the action of the Kerr-like Hamiltonian
\begin{equation*}
\overline{\bf H}_{K} =-\overline{\gamma}_K \left((\bn_a+\bn_b)^2 + 2 \bn_a \right)
\end{equation*}
for a time $t_K = \frac{\pi}{2\, \overline{\gamma}_K}$, an initial state $\q{\minou\alpha,\alpha}$ would get transformed, up to a global phase factor, into:
\begin{equation}\label{eq:EntState}
	\q{\overline{c}_{\alpha}}=(\q{\alpha,\alpha}-i\q{\minou\alpha,\minou\alpha})/\sqrt{2}\; .
\end{equation}
In the next Section, we  show that the action of $\overline{\bf H}_{K}$ can be simulated by sandwiching the resonant interaction $\bbU_r$
between two dispersive interactions. The corresponding reservoir thus stabilizes $\q{\overline{c}_{\alpha}}$.

\subsection{Composite interaction}\label{ssec:2comps}

\begin{figure}
		\begin{center}
\setlength{\unitlength}{1mm}
\begin{picture}(80,92)
\put(-1,65){\includegraphics[width=75mm]{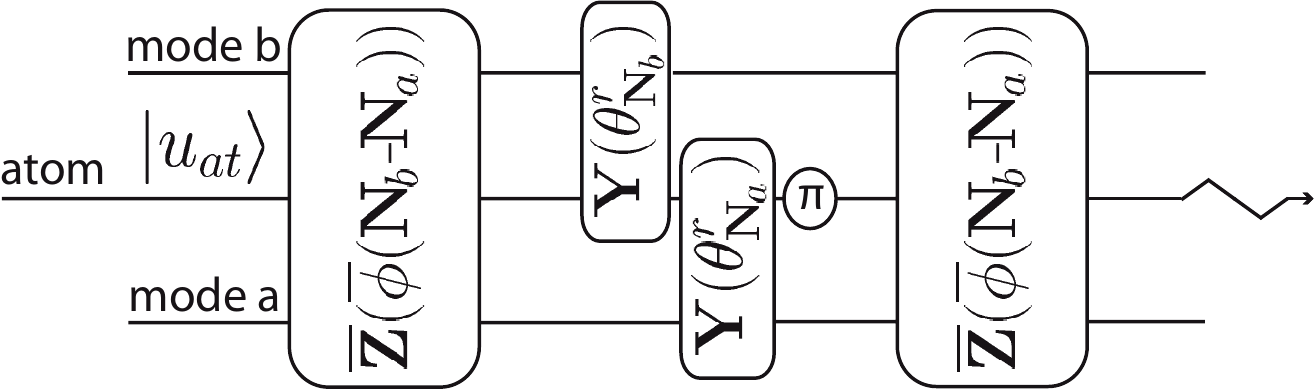}}
\put(-1,0){\includegraphics[width=75mm]{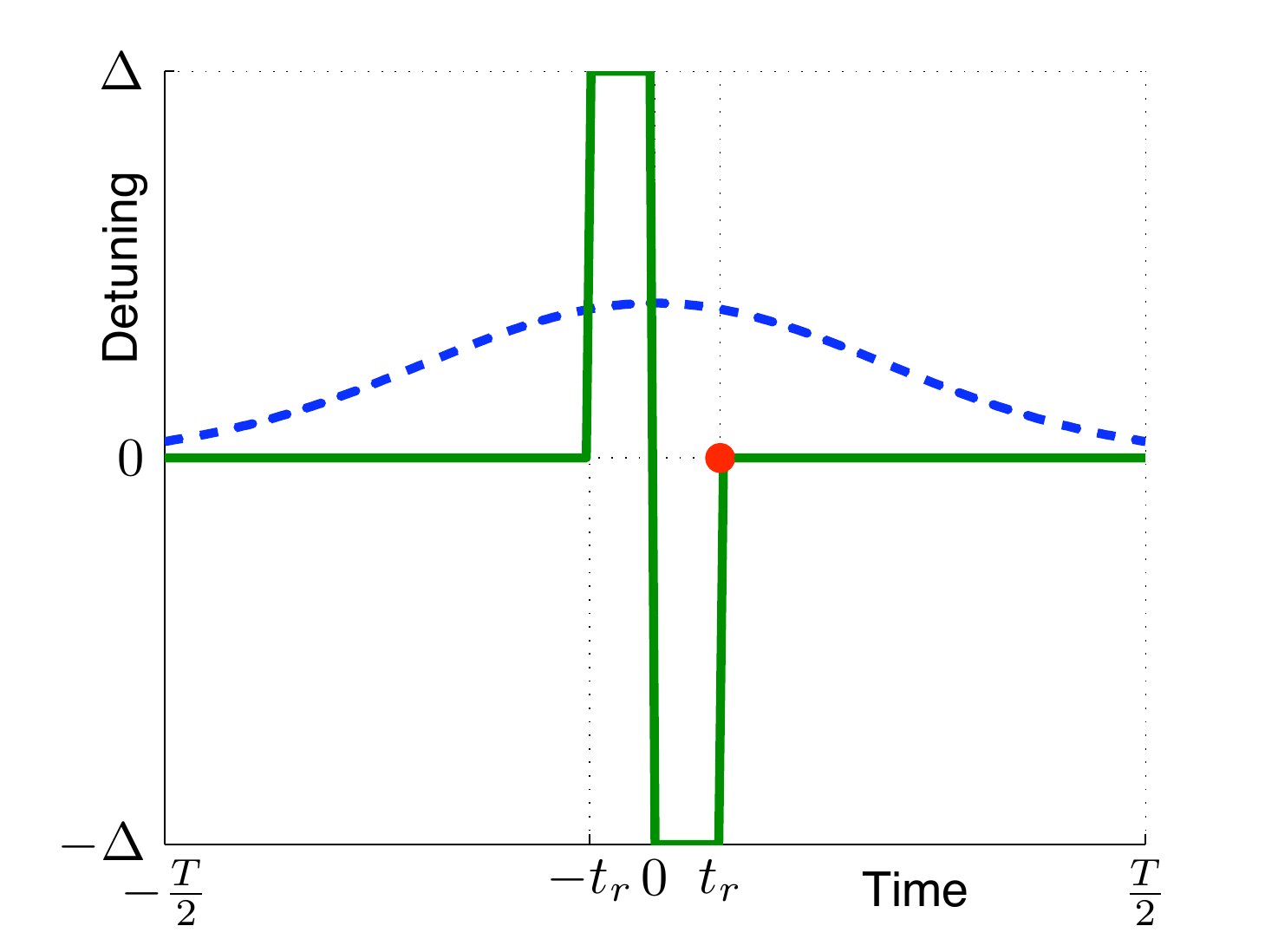}}
\end{picture}
	\end{center}
\caption{Timing of the composite interaction of the atom with the two cavity modes $a$ and $b$ at frequencies $\omega_b > \omega_a$. Bottom frame, solid line:
time profile of $\delta$ (difference between the atomic frequency $\omega_0$  and the mean frequency $\omega_m$ of the two cavity modes) during cavity crossing by one atomic sample. For $\delta = 0,\, +\Delta,\, -\Delta$ respectively, $\omega_0$ coincides with $\omega_m=\frac{\omega_b+\omega_a}{2}$, $\omega_b$, $\omega_a$. The $\pi$ pulse on the atomic state is represented here as a red dot.  Bottom frame, dashed line: coupling strength $\Omega(vt)$ with $t=0$ when the atom is at cavity center. Top frame: scheme of the propagators corresponding to the successive steps in the composite interaction.}\label{fig:timeline_2modes}
\end{figure}

The detuning profile $\delta(t)$ used to stabilize $\q{\overline{c}_{\alpha}}$ is represented on Fig.~\ref{fig:timeline_2modes} (bottom part). The atomic frequency is first set at $\omega_m$ ($\delta=0$), between $t=-T/2$ and $t=-t_r$. The atom interacts non-resonantly with both modes, with opposite detunings. We restrict in this Section to the dispersive regime. The corresponding evolution operator is $\bbZ(\overline{\phi} (\bn_b-\bn_a))$ (see Appendix~\ref{sec:propags}), describing opposite phase shifts of the two modes driven by the atom, with a phase shift per photon $\overline \phi$.

The atom is then successively set at resonance with $b$ and $a$ for a time $t_r$. During these short time intervals, we neglect the residual dispersive interaction with the other mode. The second dispersive interaction with the two modes is performed by setting again $\delta=0$  from $t_r$ to $T/2$. With this sequence, the phase shifts produced in the dispersive interactions would add up for the terms where the atom undergoes $\q{e}\qd{e}$ and $\q{g}\qd{g}$ during the resonant parts. Instead, as in the single mode case, these phase shifts must cancel out. We thus apply on the atom at $t_r$ a $\pi$ pulse on the $\q e\rightarrow \q g$ transition. It is driven by a classical source feeding a pulse of microwave with negligible duration through the interval between the cavity mirrors. This pulse does not couple into the cavity modes.

The phases of modes $a$ and $b$  evolve at the frequencies $\pm\Delta$. In order to cancel the build-up of these phases during reservoir operation, we constrain the total time $T$ between successive resonant interactions to $ T\Delta=0$ modulo $2\pi$. This condition is easily achieved with the Stark atomic tuning.

This leads, within irrelevant rotations, to the propagator (see Appendix~\ref{sec:propags} for a detailed calculation):
\begin{eqnarray}
\label{eq:Uc2modesEff}
 \bbU_{T} \approx \bbU_{\bar c}^{\text{eff}} & = & \bbZ(\overline{\phi} (\bn_b-\bn_a))\overline{\bf Y}(\theta^r_{\bn_a})\notag\\
                                &&\qquad\overline{\bf Y}(\theta^r_{\bn_b})\bbZ(\overline{\phi} (\bn_a-\bn_b))\; .
\end{eqnarray}
Setting the dispersive interactions to produce a $\overline{\phi}=\pi$ phase shift per photon, we get
$$
\bbU_{\bar c}^{\text{eff}}~=~ e^{-i t_K \overline{\bf H}_{K}} \overline{\bf Y}(\theta^r_{\bn_a})\overline{\bf Y}(\theta^r_{\bn_b}) e^{i t_K \overline{\bf H}_{K}}\; ,
$$
with $t_K \overline{\gamma}_K = \pi/2$. The resulting atomic reservoir thus stabilizes the entangled pointer state $\q{\overline{c}_\alpha}$.
Adapted detuning profiles $\delta(t)$ yield the same propagators when the interaction strength is not the same on both modes \cite{AtomCavityCoupling}.
Generalization to entangled states with more than two coherent components in each mode is straightforward, using slightly more complex detuning sequences. Indeed the latter must then be chosen to have additive instead of opposite dispersive effects on the two modes.

\subsection{Numerical simulations}\label{ssec:2simus}

We numerically solve Eq.~\eqref{eq:Uev2} and iterate the corresponding Kraus maps starting from the vacuum state with $u=\pi/4$  and $\theta_r= \pi/2$, such that the entangled field modes amplitude is of the order of 1. Decoherence is modeled as the separate coupling of each field mode with a thermal environment, with the same damping time $T_c$ and the same temperature ($n_{t}=0.05$). The interaction strength $\Omega(s)$ of the atom with each mode has the same Gaussian profile as in the single-mode case, with $\Omega_0/2\pi=50$~kHz. In the computations, the field Hilbert space is truncated to the $10$ first Fock states for each mode.

\begin{figure}[t]
		\begin{center}
\includegraphics[width=80mm]{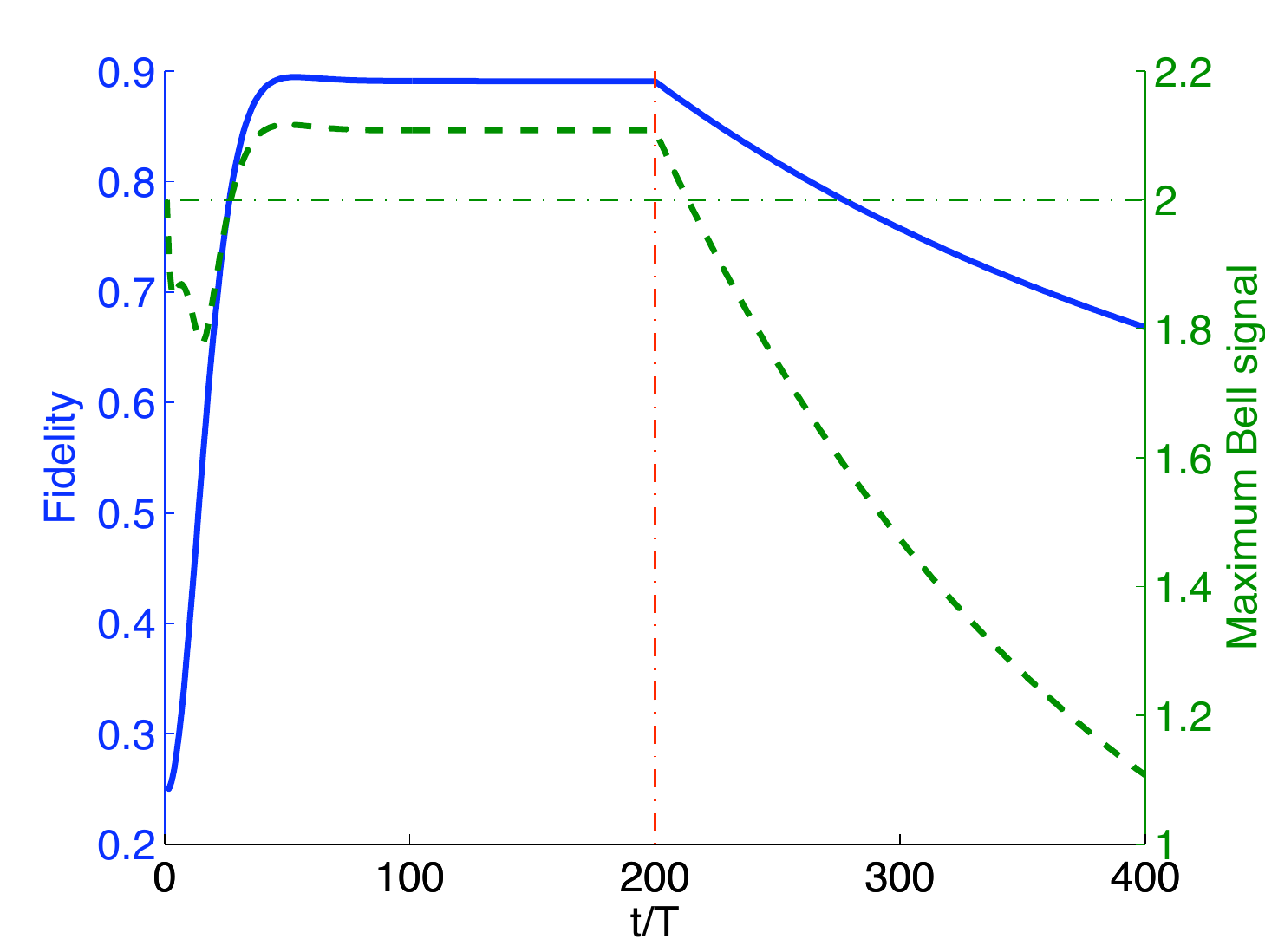}
	\end{center}
\caption{Simulation of the reservoir stabilizing a two-mode entangled state. Solid line: fidelity of $\rho$, the cavity state starting at vacuum, w.r.t.~an ideal optimized entangled state of the two modes $\q{\overline{c}_{\alpha}}$, as a function of time in units of the sample interaction time $T$. The reservoir operates up to $t/T=200$ and is then switched off. Dashed line: maximum Bell signal $\mathcal{B}^\text{max}$ as a function of time. A $\mathcal{B}^\text{max}$ value above the thin dash-dotted line ($\mathcal{B}^\text{max}=2$) proves entanglement of $\rho$.}
\label{fig:Fidelity_Bell_Realistic2ModeExp}
\end{figure}

Figure \ref{fig:Fidelity_Bell_Realistic2ModeExp} shows (solid line) the evolution of the fidelity $\qd{\overline{c}_{\alpha}} \rho \q{\overline{c}_{\alpha}}$ of the two-mode cavity state $\rho$ w.r.t.~an entangled two-component MFSS $\q{\overline{c}_{\alpha}}$ with 0.67 photons on the average, starting from the vacuum. The reference state is numerically optimized to maximize its fidelity w.r.t. the reservoir stationary state ($\approx\rho_{200}$). We have chosen $\Delta = 8\, \Omega_0$, $T_c=650$~ms. The atomic velocity is $v=22$~m/s and each atomic sample has a probability $p_{at}=0.3$ to contain one atom (we neglect here two-atom samples). The engineered reservoir is efficient, since the optimal fidelity is $\approx 89\%$. This value is reached after $\approx 30$ samples, corresponding to only 10 atoms on the average. To illustrate the protection of the state, we interrupt the reservoir after 200 atomic samples. As shown in Figure \ref{fig:Fidelity_Bell_Realistic2ModeExp}, the fidelity w.r.t. the target state rapidly decreases.

The entanglement of the state produced by the reservoir can be proved by a violation of a Bell inequality adapted to this two-mode case\cite{Banaszek-Wodkiewicz-PRL_1999,Milman-al-EurPhysJD_2005}. The Bell signal is:
\begin{eqnarray}\label{eq:BellSignal}
\mathcal{B}(\gamma_a,\gamma_b,\gamma_a',\gamma_b') & = & \tfrac{\pi^2}{4}|\overline{W}(\gamma_a',\gamma_b') + \overline{W}(\gamma_a,\gamma_b')\\
\nonumber & & \phantom{aa} + \overline{W}(\gamma_a',\gamma_b) - \overline{W}(\gamma_a,\gamma_b)|\;,	
\end{eqnarray}
where $\overline{W}(\gamma_a,\gamma_b)$ is the two-mode Wigner function. It is defined as:
$$\overline{W}(\gamma_a,\gamma_b)=\tfrac{4}{\pi^2}Tr({\bf D}^a_{-\gamma_a}{\bf D}^b_{-\gamma_b}\, \rho \, {\bf D}^a_{\gamma_a}{\bf D}^b_{\gamma_b}{\bf\overline P}) \ ,$$
where $\overline{\bf P}=e^{i\pi (\bn_a+\bn_b)}$ is a joint parity operator and ${\bf D}_{\gamma_a}^a$ and ${\bf D}_{\gamma_b}^b$ are the displacement operators for modes $a$ and $b$ respectively. In a local realistic model, $\mathcal{B}$ is always smaller than 2. A value larger than $2$ for some $(\gamma_a,\gamma_b,\gamma_a',\gamma_b')$ amplitudes is a proof that $\rho$ is not separable.

\begin{figure}[t]
		\begin{center}
\includegraphics[width=80mm]{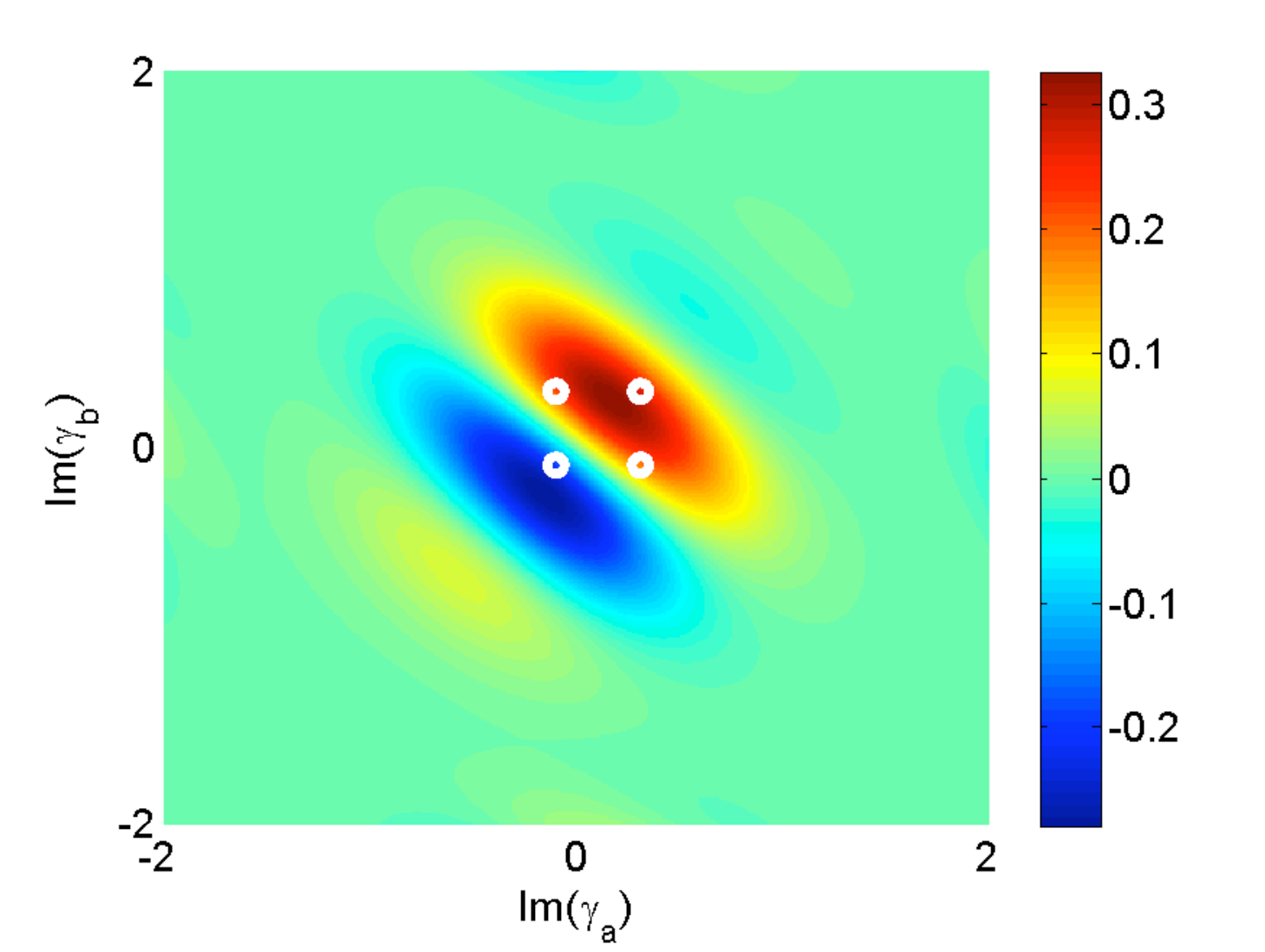}
	\end{center}
\caption{Cut in the plane ($\Re(\gamma_a)=\Re(\gamma_b)=0$) of the two-mode Wigner function $\overline{W}(\gamma_a,\gamma_b)$ of $\rho_{200}$. The fringes and negative values for $\overline W$ are a signature of the ``quantumness'' of the stabilized state. The white dots show the points used to maximize the violation of Bell's inequality.}
\label{fig:WignerFunction_2ModeExp}
\end{figure}

Figure \ref{fig:WignerFunction_2ModeExp} shows a cut of the two-mode Wigner function of $\rho_{200}$ in the plane $\Re(\gamma_a)=\Re(\gamma_b)=0$ in which maximum violation of the inequality is expected \cite{Banaszek-Wodkiewicz-PRL_1999}. A numerical optimization of the Bell signal in this plane provides the four amplitudes shown as white dots. We have performed similar optimizations of $\mathcal{B}$ after each atomic sample interaction and plotted the maximum Bell signal $\mathcal{B}^\text{max}$ as a dashed line in figure  \ref{fig:Fidelity_Bell_Realistic2ModeExp}. It reaches $\approx2.1>2$ which implies that the reservoir stabilizes a provably entangled state of the modes. When the reservoir is switched off after 200 interactions, decoherence causes a rapid decrease of $\mathcal{B}^\text{max}$.

\begin{figure}[!t]
		\begin{center}
{\includegraphics[width=80mm]{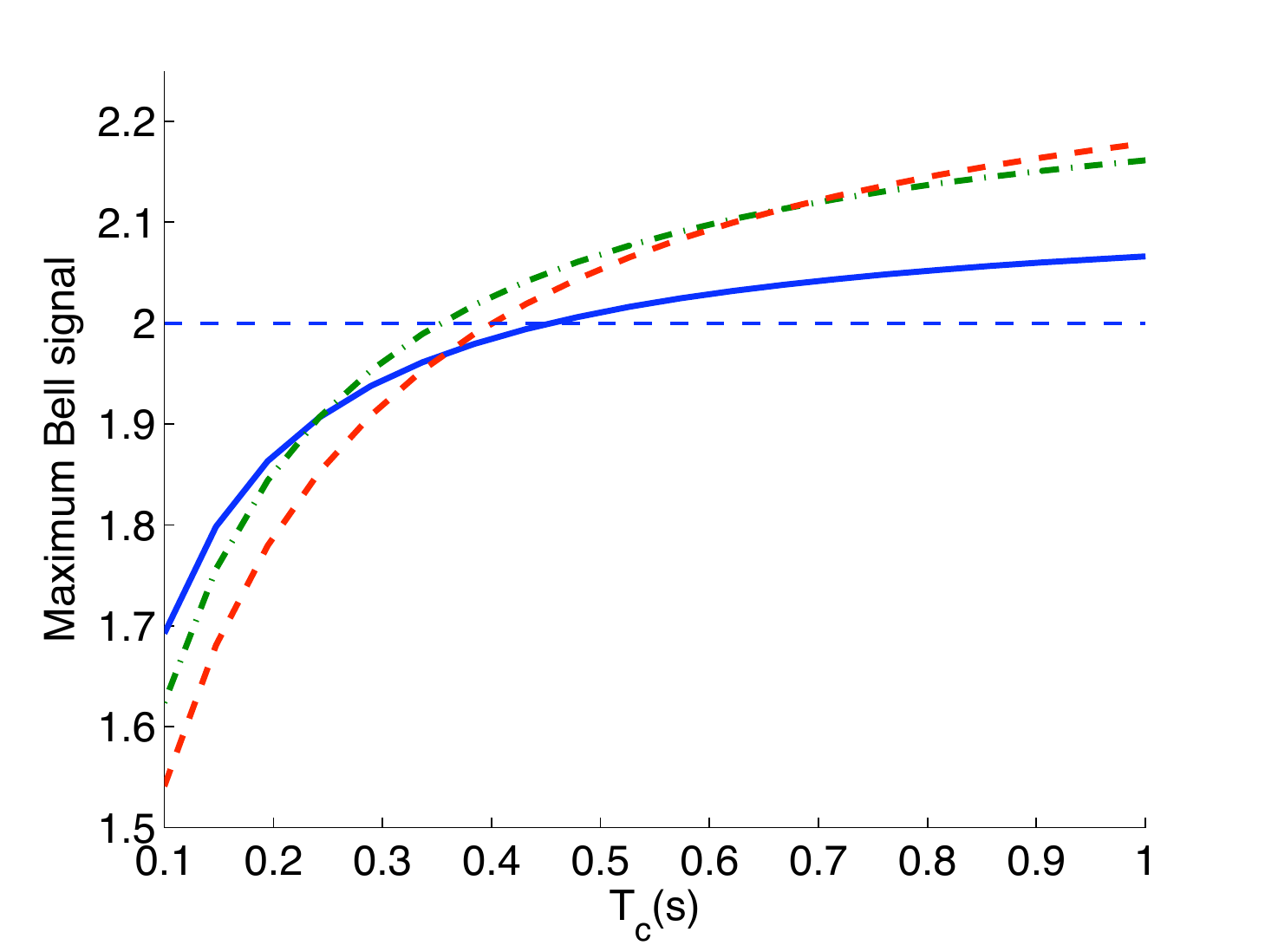}}
	\end{center}
\caption{Maximum Bell signal $\mathcal{B}^\text{max}$ of $\rho_{200}$ as a function of the cavity lifetime $T_c$ for $\Delta/2\pi=300$~kHz, $v=30$~m/s (solid blue line); $\Delta/2\pi=400$~kHz, $v=22$~m/s (dashed-dotted green line); $\Delta/2\pi=500$~kHz, $v=18$~m/s (dashed red line). }
\label{fig:BellTcav}
\end{figure}

Figure \ref{fig:BellTcav} shows the maximum Bell signal $\mathcal{B}^\text{max}$ of the steady state as a function of $T_c$, for three detuning and atomic velocity values. The Bell inequality is violated for all these settings when $T_c>450$~ms. The crossing of the different curves illustrates the competition between two effects. For small $T_c$s, the Bell signal is larger when $\Delta$ is smaller, since a small $\Delta$ corresponds to a relatively large velocity and thus to a smaller total interaction duration $T$. Thus the reservoir is a more efficient protection against decoherence when $\Delta$ is small.
For very large $T_c$, cavity damping becomes negligible w.r.t.~the dispersive approximation error introduced in the reservoir action, for which large $\Delta$ values are preferred.

The $T_c$ values required for a violation are certainly difficult to reach, but they are only $\approx 3$ times larger than the best damping time reported so far~\cite{Kuhr-al-APL_2007}. Stabilizing field states violating a Bell inequality may thus be within reach of the next generation of experiments.

\section{Concluding remarks}\label{sec:Conclusion}

We have proposed simple engineered reservoirs stabilizing a wide variety of non-classical field states in one and two quantum cavity modes. These reservoirs efficiently counteract the standard relaxation of the cavities and offer promising perspectives for studies and applications of mesoscopic field state superpositions.

We have gained a detailed insight into the reservoir mechanisms, and particularly into the way it corrects for decoherence-induced quantum jumps of the field. We have performed extensive numerical simulations justifying the approximations used in~\cite{PRancestor} and assessing the robustness of the method to experimental imperfections.

We have discussed here, for the sake of definiteness, the reservoir operation in the context of the microwave-CQED experiments performed with circular Rydberg atoms and superconducting cavities at ENS. We have shown that many quantum states protected by our reservoir could realistically be observed in this context. Clearly, the method could be straightforwardly extended to other spin/spring systems, in cavity QED and trapped ions contexts. It is particularly appealing for the thriving field of circuit-QED~\cite{Devoret-Martinis-QInfoProcess_2004}. Resettable superconducting qubits~\cite{Reed-al-Schoelkopf-APL_2010} interacting with one or two cavity modes could be used to implement our proposal. With two separate cavities interacting with qubits, it would become possible to stabilize a non-local entangled mesoscopic superposition and to study the fascinating interplay between decoherence and non-locality.


\appendix
\section{Propagators}
\label{sec:propags}

This appendix details the computation of the propagators associated to the atom-cavity interaction in the various settings used in the main text.

\subsection{Single-mode case}

For a resonant interaction [$\delta(t) =0$], Eq.~(\ref{eq:Uev}) writes:
$$\tfrac{d}{dt} {\bf U}(t) = \frac{\Omega(s)}{2} (\, \vert g \rangle \langle e \vert\, \text{\bf{a}}^{\dagger} - \vert e \rangle \langle g \vert\,  \text{\bf{a}} \, ) \;  {\bf U}(t)\ ,$$
with $s=vt$ if we set the time origin such that the atom crosses the cavity axis at $t=0$. For each Bloch sphere $B_n$ associated to the invariant space spanned by $(\q{g,n+1},\q{e,n})$, this interaction induces a Rabi rotation at an angular rate $\sqrt{n} \Omega(s)$ around the $Y$ axis. We therefore define the unitary operator ${\bf Y}(f_\bn)$ [Eq.~\eqref{eq:Y}] performing a rotation around $Y$ by an angle $f(n)$, where $f(n)$ is an arbitrary function of $n$. The resonant interaction propagator is thus given by Eq.~\eqref{eq:Uresonant}.

For the interaction between $t_1$ and $t_2$ with a constant nonzero detuning $\delta(t) = \overline{\delta} \neq 0$~\cite{RemDelta}, the Gaussian variation of $\Omega(vt)$ precludes an exact integration of Eq.~(\ref{eq:Uev}). However, assuming that $\Omega(vt)$ varies slowly enough, the coupled atom-field system evolves adiabatically. An initial eigenstate of ${\bf H}_{JC}(t_1)$ (a ``dressed state") then remains, for any time $t$, close to an eigenstate of ${\bf H}_{JC}(t)$~\cite{HarocheBook}. This adiabatic approximation is valid provided:
\begin{equation}\label{eq:AdCond}
\left| \tfrac{2 v}{w\Omega_0\sqrt{n+1}} s e^{-s^2} \right| \ll \left(\tfrac{\overline{\delta}}{\Omega_0\sqrt{n+1}} \right)^2 + e^{-2s^2},\quad \forall s \in (\tfrac{t_1 v}{w},\tfrac{t_2 v}{w})\;,
\end{equation}
for all $n$ in the relevant photon number range.

The dressed states $(\q{-,n}_t,\q{+,n}_t)$ that diagonalize ${\bf H}_{JC}(t)$ for each $n=1,2,...$ satisfy
$${\bf H}_{JC}(t) \, \q{\pm,n}_t = \pm \frac{\overline{\delta}}{2} \sqrt{1+ (n+1)\left(\frac{\Omega(vt)}{\overline{\delta}}\right)^2} \; \q{\pm,n}_t\ , $$
and write explicitly
\begin{align}
\q{-,n}_t &= \cos (\xi_n^{(t)}/2) \,\q{g,n+1}+ i \sin (\xi_n^{(t)}/2) \,\q{e,n}  \notag \\
\q{+,n}_t &= i\sin(\xi_n^{(t)}/2) \,\q{g,n+1} + \cos (\xi_n^{(t)}/2) \,\q{e,n}\ ,   \label{eq:td:HjcEigVecs}
\end{align}
where we define $\xi_n^{(t)}$ by
\begin{equation}\label{eq:xi}
\tan \xi_n^{(t)} = \tfrac{\Omega(vt/2) \sqrt{n}}{\overline{\delta}} \quad \text{with } \xi_n^{(t)} \in (\tfrac{-\pi}{2},\tfrac{\pi}{2}) \; .
\end{equation}

The propagator ${\bf U}_q$ corresponding to the parameter set $q=(t_1,t_2,v,\overline{\delta})$ is thus
\begin{multline}
{\bf U}_q=\sum_n \q{-,n}_{t_2} \qd{-,n}_{t_1} \; e^{\frac{i}{2}\phi^q_{n+1}}\\
+\;  \q{+,n}_{t_2} \qd{+,n}_{t_1} \; e^{\frac{-i}{2}\phi^q_{n+1}}\;,
\label{eq:lanouvelleforme}
\end{multline}
where the accumulated phase $\phi^q_{n}$ is given by:
\begin{equation}\label{eq:phi}
\phi^q_n = \overline{\delta} \; \int_{t_1}^{t_2} \,  \sqrt{1+ n ( \Omega(vt)/\overline{\delta})^2 } \, dt \; .
\end{equation}

The restriction of ${\bf U}_q$ on the Bloch sphere $B_n$ can then be written as:
\begin{eqnarray}
\nonumber	& & \!\!\!\!\!\!\!\!\!\!\q{-,n}_{t_2} \qd{-,n}_{t_1} \; e^{\frac{i}{2}\phi^q_{n+1}} \;+\;  \q{+,n}_{t_2} \qd{+,n}_{t_1} \; e^{\frac{-i}{2}\phi^q_{n+1}} \\[2mm]
\nonumber  & = & (\q{-,n}_{t_2}\qd{g,n+1} \;+\; \q{+,n}_{t_2}\qd{e,n}) \\
\nonumber & & \times(\q{g,n+1}\qd{g,n+1}\; e^{\frac{i}{2}\phi^q_{n+1}} \;+\; \q{e,n}\qd{e,n}\; e^{\frac{-i}{2}\phi^q_{n+1}}) \\
\label{eq:ZeChange}  & & \times(\q{-,n}_{t_1}\qd{g,n+1} \;+\; \q{+,n}_{t_1}\qd{e,n})^\dagger\ .
\end{eqnarray}

The transformation $(\q{-,n}_{t} \qd{g,n+1} \;+\; \q{+,n}_{t} \qd{e,n})$ is a rotation around the $X$-axis of $B_n$ by an angle $-\xi^{(t)}_{n+1}$. The transformation $(\q{g,n+1}\qd{g,n+1}\; e^{i\phi^q_{n+1}/2} + \q{e,n}\qd{e,n}\; e^{-i\phi^q_{n+1}/2})$ is a rotation around the $Z$-axis of $B_n$ by an angle $\phi^q_{n+1}$. We thus introduce in Eqs.~\eqref{eq:X},\eqref{eq:Z} the unitary operators ${\bf X}(f_\bn)$ and ${\bf Z}(f_\bn)$ representing these rotations on each Bloch sphere $B_n$. Noting that ${\bf X}(-f_\bn)^\dagger ={\bf X}(f_\bn)$, we can finally write \eqref{eq:lanouvelleforme} in the compact form:
\begin{equation}\label{eq:Uadiabatic}
	{\bf U}_q = {\bf X}(-\xi^{(t_2)}_{\bn})\; {\bf Z}(\phi^q_{\bn})\; {\bf X}(\xi^{(t_1)}_{\bn}) \; .
\end{equation}
At the start and end of the complete composite interaction, the atom-cavity coupling is weak: $\Omega^2(\pm vT/2) = \Omega_0^2/100$. We can thus take ${\bf X}(-\xi^{(-T/2)}_{\bn}) = {\bf X}(\xi^{(T/2)}_{\bn}) = \bid$ in Section \ref{sec:ArbitDetun} since $\xi_\bn^{(\pm T/2)}\, \approx\, 0$. This leads to Eq.~\eqref{eq:UcOps}.

In the large detuning regime studied in Section~\ref{sec:LargeDetun}, we can even neglect all the ${\bf X}$ operators in ${\bf U}_q$ compared to the large dispersive phase shift operator ${\bf Z}(\phi^q_{\bn})$.

\subsection{Two-mode case}
\label{ssec:app2modesUc}

In the two-mode case, it is not possible to get an exact expression for the dressed states. We thus restrict either to a resonant interaction with one of the modes or to a dispersive interaction with both, assuming a large detuning $2\Delta$ between modes $a$ and $b$. In the resonant case, we neglect the residual dispersive interaction with the other mode. For the non-resonant interaction, we use simple first-order dispersive expressions. In both cases, simulations  integrating Eq.~\eqref{eq:Uev2} explicitly confirm the validity of our approximations.

Let us first investigate the resonant case, with $\delta = \pm \Delta$. A simple adaptation of the single mode results leads to:
\begin{eqnarray}\label{eq:a2:res}
\nonumber\bbU_q & = &  e^{- i \Delta (\bn_b-\bn_a) (t_2-t_1)} \, \overline{\bf Z}(\Delta(t_2-t_1)) \, \overline{\bf Y}(\theta^q_{\bn_b})\\
& & \text{for } q=(t_1,t_2,v,\Delta) \\
\label{eq:a2:res2}
\nonumber \bbU_q & = & e^{- i \Delta (\bn_b-\bn_a) (t_2-t_1)} \, \overline{\bf Z}(\Delta(t_1-t_2)) \, \overline{\bf Y}(\theta^q_{\bn_a})\\
& & \text{for } q=(t_1,t_2,v,-\Delta)\; ,
\end{eqnarray}
where $\overline{\bf Y}(\theta^q_{\bn_{a}})$, for instance, is the tensor product of ${\bf Y}(\theta^q_{\bn_a})$ acting on the pair atom-mode $a$, with the identity acting on $b$. We define a generalized two-mode phase rotation by: \begin{equation}\label{eq:bbZ}
	\bbZ(f_{\bn_a,\bn_b}) = \q{g}\qd{g}\, e^{\tfrac{i}{2}\,f_{\bn_a,\bn_b}} + \q{e}\qd{e}\, e^{\tfrac{-i}{2}\,f_{(\bn_a+\bid),(\bn_b+\bid)}}\ ,
\end{equation}
where the operator $f_{\bn_a,\bn_b}$ is diagonal in the joint Fock state basis of the two modes with $f_{\bn_a,\bn_b}\, \q{n_a,n_b} = f(n_a,n_b) \q{n_a,n_b}$. In Eqs.~\eqref{eq:a2:res} and \eqref{eq:a2:res2}, $\overline {\bf Z}$ is used with a constant argument $f_{\bn_a,\bn_b}=\pm\Delta(t_2-t_1)$.

We consider now the dispersive interaction corresponding here to $\delta=0$. Applying second-order perturbation theory in $\Omega_0/\Delta$, we get for $q=(t_1,t_2,v,0)$:
\begin{eqnarray}\label{eq:a2:disp}
\bbU_q & = & e^{- i \Delta (\bn_b-\bn_a) (t_2-t_1)} \, \bbZ(\overline{\phi}^q(\bn_b-\bn_a))\;,
\end{eqnarray}
with $\overline{\phi}^q = \frac{1}{2\Delta} \, \int_{t_1}^{t_2} \Omega^2(vt) \, dt \; .$

Using Eqs. \eqref{eq:a2:res},\eqref{eq:a2:res2},\eqref{eq:a2:disp} and the commutation relation \eqref{eq:NcommA}, we get an approximate evolution operator with the sequence defined in Section~\ref{sec:2modes} (with $T\Delta=0$ modulo $2\pi$):
\begin{eqnarray}
\label{eq:Uc2modes}
\bbU_{T} \approx \bbU_{\bar c}^{\text{eff}} & = &  \bbU_\pi \; \overline{\bf Z}(-\Delta (T/2+t_r))\notag \\
	              & & \bbZ(\overline{\phi} (\bn_b-\bn_a)) \; \overline{\bf Y}(\theta^r_{\bn_a})\; \overline{\bf Y}(\theta^r_{\bn_b})\;\;\;\phantom{karam} \\
	              & & \bbZ(\overline{\phi} (\bn_a-\bn_b)) \; \overline{\bf Z}(\minou\Delta (T/2-t_r))\notag\;,
\end{eqnarray}
with
\begin{eqnarray}
\label{eq:Uc2params} \overline{\phi} & = & \frac{1}{2\Delta} \, \int_{-T/2}^{-t_r} \, \Omega^2(vt) \, dt \, .
\end{eqnarray}

The first line in Eq.~\eqref{eq:Uc2modes} has no effect on the Kraus map since it is a rotation on the atom only after it has interacted with the modes. The operator $\overline{\bf Z}(-\Delta (T/2-t_r))$ can simply be compensated by properly setting the phase of the Ramsey pulse, preparing now each atom in $\overline{\bf Z}(\Delta (T/2-t_r))\q{u_{at}}$.  These considerations lead to the effective propagator given in Eq. \eqref{eq:Uc2modesEff}.


\section{Equilibrium of reservoir with damping}
\label{sec:dampeq}

If  $\rhoK_\infty$ of the form~\eqref{eq:rhoKinfty} is a stationary solution of~\eqref{eq:rhoKDecoherence} then we have:
\begin{multline*}
    \int_{-\alpha^c_\infty}^{\alpha^c_\infty}
    \mu(z) \left(\beta-\tfrac{\kappa+\kappa_c}{2} z \right) \big( (\ba^\dag -z) \q{z}\qd{z} + \q{z}\qd{z}  (\ba -z) \big)~dz
    \\
    + \int_{-\alpha^c_\infty}^{\alpha^c_\infty}
    \kappa_c (\mu(-z)-\mu(z)) z^2  \q{z}\qd{z} ~dz =0\ ,
\end{multline*}
(using  $\ba \q{z} = z \q{z}$, $e^{i\pi\bn} \ba \q{z} = z\q{\minou z}$ and their Hermitian conjugates).
For any real $\xi$, multiplying on the left by coherent state $\qd{\xi}$ and on the right by $\q{\xi}$ yields
$$
    \int_{-\alpha^c_\infty}^{\alpha^c_\infty}
    2 \mu(z) \left(\beta-\tfrac{\kappa+\kappa_c}{2} z \right) (\xi -z) e^{-(\xi-z)^2}~dz
    + \int_{-\alpha^c_\infty}^{\alpha^c_\infty}
    \kappa_c  z^2 (\mu(-z)-\mu(z))  e^{-(\xi-z)^2} ~dz =0\ ,
$$
since $|\bket{\xi|z}|^2= e^{-(\xi-z)^2}$, $\xi$ and $z$ being real. An integration by parts of the first integral yields
\begin{multline*}
\left[\mu(z) \left(\beta-\tfrac{\kappa+\kappa_c}{2} z \right) e^{-(\xi-z)^2}\right]_{z=-\alpha_\infty^c}^{z=\alpha_\infty^c}
- \int_{-\alpha^c_\infty}^{\alpha^c_\infty}
    \bigg(\tfrac{d}{dz} \big(\mu(z) \left(\beta-\tfrac{\kappa+\kappa_c}{2} z \right) \big)\bigg)e^{-(\xi-z)^2} ~dz
    \\  + \int_{-\alpha^c_\infty}^{\alpha^c_\infty}
    \kappa_c  z^2 (\mu(-z)-\mu(z))  e^{-(\xi-z)^2} ~dz  = 0 \, .
\end{multline*}
Since this holds for any $\xi$ real, the only possibility is
$$\kappa_c z^2 (\mu(-z)-\mu(z)) - \tfrac{d}{dz} \big(\mu(z) \left(\beta-\tfrac{\kappa+\kappa_c}{2} z \right) \big)=0$$
for $z\in(-\alpha_\infty^c,\alpha_\infty^c)$ with the boundary conditions
$\lim_{z\mapsto \alpha_\infty^c} \mu(z)(z-\alpha^c_\infty)= 0$ and $\mu(-\alpha_\infty^c)=0 \, .$

To solve this differential equation for  $z\in [-\alpha^c_\infty,\alpha^c_\infty]$, we decompose $\mu$ in its even and odd parts: these parts satisfy two first-order coupled differential equations that can be integrated directly to give formula \eqref{eq:mu} for $\mu(z)$.


\section*{Acknowledgements}

The authors thank S. Haroche, M. Mirrahimi and I. Dotsenko for enlightening discussions and references. AS has been a FNRS postdoctoral fellow at U.~Li\`ege, visiting researcher at Mines ParisTech and member of the IAP network DYSCO. ZL acknowledges support from Agence Nationale
de la Recherche (ANR), Projet Jeunes Chercheurs EPOQ2 number ANR-09-JCJC-0070. JMR and MB acknowledge support from the EU and ERC (AQUTE and DECLIC projects). The authors were partially supported by the ANR, Projet Blanc EMAQS ANR-2011-BS01-017-01, Projet Blanc QUSCO-INCA ANR-09-BLAN-0123 and Projet C-QUID BLAN-3-139579.



\bibliographystyle{plain}

\begin{thebibliography}{99}
\bibitem{FNcoh}
The components are well separated only in the case $k < \pi \alpha$~\cite[Section~7.2]{HarocheBook}.
\bibitem{PierreApprox}
Explicitly, we have (all $\q{\cdot}$ denote coherent states)
\begin{eqnarray*}
  \epsilon\bn \q{\beta} &=&
     \q{(1+\epsilon)\beta} - \q{\beta} + |\beta|^2 \Re(\epsilon) \q{\beta}
     \\
  \epsilon \ba^\dag \q{\beta} &=& \q{\beta+\epsilon} - \q{\beta} + \Re(\epsilon \beta\daag)\q{\beta}
  \\
  \sum_\nu a_\nu \q{\beta+\epsilon_\nu} &=&
  \left(\sum_\nu a_\nu \right) \q{\tfrac{\sum_\nu a_\nu (\beta+\epsilon_\nu)}{\sum_\nu a_\nu}}\;,
\end{eqnarray*}
up to second order terms in $|\epsilon|$, $|\epsilon_\nu|$, with $\epsilon,\epsilon_\nu,a_\nu\in\mathbb{C}$ and $\sum_\nu a_\nu \neq 0$.
\bibitem{FNchi} Note that for all $n>0$, phase $\chi^c_n$ is well-defined because $\vert \xi_n \vert \in (0, \pi/2)$. For $n=0$, the term multiplying the undefined $\chi^c_0$ vanishes in Eq.~\eqref{eq:exactUt}.
\bibitem{AtomCavityCoupling}
Consider interaction strength $\Omega_a(s)=\Omega(s)$ with mode $a$ and $\Omega_b(s)=\lambda \Omega(s)$ with mode $b$, where $\lambda \neq 1$. Adjusting the resonant interaction times to get the same coherent state amplitude in both modes is straightforward. Equal dispersive phases are obtained by setting $\delta(t) \approx \bar\delta_0=\Delta\frac{1-\lambda^2}{1+\lambda^2} \neq 0$ for $t \in (-T/2,-t_r/2) \cup (t_r/2,T/2)$. Indeed, to first order approximation in $\Omega_0/\Delta$, this choice shifts the eigenvalues of the Hamiltonian by equal amounts $\Omega_a(s) /\sqrt{1+\frac{\bar\delta_0}{\Delta}} = \Omega_b(s)/\sqrt{1-\frac{\bar\delta_0}{\Delta}}$ for the two modes.
\bibitem{RemDelta} The following developments also carry through with (slowly) continuously varying $\delta(t)$. For pedagogical reasons, we have chosen to explain the experiment with piecewise constant $\delta$. We therefore directly particularize the notation to $\delta(t)=\overline{\delta}$ constant.
\end{thebibliography}

\end{document}